\documentclass{emulateapj}
\usepackage{apjfonts}
\pdfoutput=1
\makeatletter

\newcommand{\bc}{\begin{center}}
\newcommand{\ec}{\end{center}}
\newcommand{\be}{\begin{equation}}
\newcommand{\ee}{\end{equation}}
\newcommand{\ba}{\begin{eqnarray}}
\newcommand{\ea}{\end{eqnarray}}
\newcommand{\bt}{\begin{tabular}}
\newcommand{\et}{\end{tabular}}

\newcommand{\nh}{n_{\rm H}}

\def\farcs{\hbox{$.\!\!^{\prime\prime}$}}

\newcommand{\chan}{{\sl Chandra}}

\newcommand{\edot}{\dot{E}}

\def\aciss3{{ACIS-S3}}

\def\edot{\dot{E}}

\def\ls{\lower 2pt \hbox{$\;\scriptscriptstyle \buildrel<\over\sim\;$}}

\slugcomment{}
\shorttitle{X-ray observations of parsec-scale pulsar tails}

\makeatother
\begin{document}

\title{X-ray observations of parsec-scale  tails
behind two middle-aged pulsars}

\author{ O.\ Kargaltsev, Z.\ Misanovic, G.\ G.\ Pavlov,  J.\ A.\ Wong,
 \& G.\ P.\ Garmire}

\affil{Dept. of Astronomy and Astrophysics,
The Pennsylvania State
University, 525 Davey Lab. University Park,
PA 16802}

\begin{abstract}

 {\sl Chandra} and {\sl
XMM-Newton} resolved extremely long tails behind two middle-aged
pulsars, J1509--5850 and J1740+1000.
The tail of PSR J1509--5850 is discernible
up to $5.6'$ from
the pulsar, which corresponds to the projected length
$l_\perp =6.5d_4$ pc, where  $d=4 d_4$ kpc is the distance to the pulsar.
The observed tail flux is
$2\times10^{-13}$ erg s$^{-1}$ cm$^{-2}$ in the 0.5--8 keV
band. The tail spectrum fits an absorbed power-law
(PL) model with the photon index $\Gamma=2.3\pm0.2$, corresponding to the
0.5--8 keV luminosity of
$1\times 10^{33}d_4^2$ ergs s$^{-1}$,
for
$n_{\rm H}= 2.1\times 10^{22}$ cm$^{-2}$.
The tail of
PSR J1740+1000
is firmly
detected up to $5'$
($l_\perp\sim 2d_{1.4}$ pc),
with a flux of
$6\times10^{-14}$  ergs cm$^{-2}$ s$^{-1}$ in the 0.4--10 keV band.
The PL fit to the spectrum
 measured from a brighter, $3'$-long,
  portion of the tail
yields $\Gamma = 1.4$--1.5
and $\nh \approx 1\times 10^{21}$ cm$^{-2}$; its
0.4--10 keV luminosity is
$\sim 2\times10^{31}d_{1.4}^2$\,ergs s$^{-1}$.
The luminosity of the entire tail is likely
a factor of
 3--4 higher.
  The large
extent  of the tails
  suggests that the bulk flow in the tails starts as mildly
relativistic downstream of the termination shock,
  and then gradually decelerates.
Within the observed extent of the J1509--5850 tail,
the average flow speed exceeds 5,000 km s$^{-1}$, and
the equipartition magnetic field  is a few $\times 10^{-5}$ G.
For the J1740+1000 tail, the equipartition field is
a factor of a few lower.
 The harder spectrum of the J1740+1000 tail
implies  either less efficient cooling
or a harder
spectrum
  of injected electrons.
 For the high-latitude PSR J1740+1000,
the
orientation of the tail on the sky
  shows
 that the pulsar is moving
 toward the Galactic plane, which means that
it was born from a halo-star progenitor.
 The comparison between the J1509 and J1740 tails
and the X-ray tails of other
pulsars shows
that the X-ray radiation efficiency
correlates poorly with the
pulsar
spin-down luminosity or age.
The X-ray efficiencies  of the ram-pressure
confined pulsar wind nebulae
(PWNe)
are systematically higher
  than those of PWNe around slowly moving pulsars with
similar spin-down parameters.

\end{abstract}


\section{Introduction}

 The multiwavelength synchrotron emission from ultrarelativistic
particles produced in the pulsar magnetosphere and
shocked
in the ambient medium is observed as a pulsar-wind nebula (PWN;
Rees \& Gunn 1974; Kennel \& Coroniti 1994;
Arons 2007; Kirk \& Lyubarsky 2007).
 Thanks to the high sensitivity and angular resolution of the {\sl Chandra}
and {\sl XMM-Newton} observatories, about 50 PWNe have been detected in X-rays
 (Kaspi et al.\ 2006; Gaensler \& Slane 2006; Kargaltsev \& Pavlov
2008, hereafter KP08).
The X-ray observations have shown that PWNe
 have complex morphologies, including toroidal structures
around the
pulsar,  jets along the pulsar's spin axis, and cometary tails (see KP08
for  {\sl Chandra} images).
 In particular, if a pulsar moves with a supersonic speed,
$v\gg c_s$, the ram
pressure, $p_{\rm ram} = \rho_{\rm amb} v^2$,
exceeds the ambient
medium pressure, $p_{\rm amb}$, resulting in a
 bowshock PWN with a  tail behind the pulsar. The well-known
 example of
such a {\em bowshock-tail} PWN is ``the Mouse PWN''
produced by the young, energetic pulsar J1747$-$2958
($\tau=26$ kyr, $\edot=2.5\times 10^{36}$ ergs s$^{-1}$),
which shows an X-ray tail of the projected length $l_\perp \sim 1$ pc
(Gaensler et al.\ 2004).
In addition to the Mouse, about
a dozen X-ray PWNe with bowshock-tail morphologies
have been discovered recently (KP08).
Interestingly, one of the longest tails,
$l_\perp \sim 1.5$ pc, was found
 behind the 3 Myr old  pulsar
  B1929+10, with a relatively low spin-down power,
$\edot=3.8\times 10^{33}$ ergs s$^{-1}$
 (Wang et al.\ 1993; Becker et al.\ 2006; Misanovic et al.\ 2007).
Although most of
 the known PWNe are powered by much younger and more powerful
 pulsars,
  it has been noticed  (Kargaltsev et al.\ 2007) that
 bowshock
 PWNe with extended tails are generally  brighter than PWNe around
 slowly moving pulsars.

\begin{deluxetable*}{lll}
\tablewidth{0pt}
\setlength{\tabcolsep}{0.07in} \tablecaption{
Observed and derived pulsar parameters}
\tablehead{\colhead{Parameter} & \colhead{J1509--5850} & \colhead{J1740+1000} } 
\startdata
R.A.\ (J2000) \dotfill
& $15^{\rm h}09^{\rm m}27\fs13(3)$ & $17^{\rm h}40^{\rm m}25\fs950(5)$ \\
Decl.\ (J2000) \dotfill
& $\!\!\!\!\!$$-58^\circ 50' 56\farcs1(5)$ &
$\!\!\!\!\!$$+10^\circ 00' 06\farcs3(2)$ \\
Epoch of position (MJD) \dotfill
& 51,463 & 51,662 \\
Galactic longitude \dotfill
  & $319.97^\circ$ & $34.01^\circ$ \\
 Galactic latitude \dotfill
& $\!\!\!\!\!$$-0.62^\circ$ & $20.27^\circ$ \\
Spin period, $P$ (ms) \dotfill
&  88.9 &  154.1 \\
Period derivative, $\dot{P}$ ($10^{-14}$) \dotfill
& 0.92 & 2.15 \\
Dispersion measure, DM (cm$^{-3}$~pc)\dotfill  & 137.7
&  23.85 \\
Distance\tablenotemark{a},
$d$ (kpc) \dotfill
& 3.8
& 1.4
\\
Distance from the Galactic plane,
$z$ (kpc) \dotfill  & 0.04
& 0.48
\\
Surface magnetic field, $B_s$ ($10^{12}$ G) \dotfill & 0.91 &  1.84 \\
Spin-down power, $\dot{E}$~($10^{35}$ erg s$^{-1}$) \dotfill
& 5.1 &  2.3 \\
Spin-down age, $\tau=P/(2\dot{P})$, (kyr) \dotfill
& 154 & 114  \\
\enddata
\tablecomments{Based on the data from Kramer et al.\ (2003)
and McLaughlin et al.\ (2002).
Figures in parentheses represent $1 \sigma$ uncertainties in least-significant
digits quoted.}
\tablenotetext{a}{
The distances are based on the dispersion measure
and the Galactic electron density distribution model by Taylor \&
 Cordes (1993).
The Cordes \& Lazio (2002) model gives $d=2.6$ and 1.2 kpc for J1509 and
J1740, respectively. }
\end{deluxetable*}

The shape of the shock, the length of the tail, and the overall
appearance of the entire bowshock-tail PWN
are expected to depend on
the interplay between
$\dot{E}$, $p_{\rm amb}$, $\rho_{\rm amb}$, and $v$,
as well as on
 the wind magnetization parameter $\sigma$ and the angle between
 the pulsar's
spin axis and the direction  of the pulsar's motion.
 Analytical and
 numerical modeling of magnetized outflows in
 bowshock-tail PWNe has been done by
  Romanova et al.\ (2005)  and Bucciantini et al.\ (2005; B05
hereafter), respectively.
 It was assumed in these works that
the magnetic field in the tail
is purely
 toroidal, and the pulsar's velocity is aligned with the
rotation axis. In addition,
  the numerical  calculations by B05
assumed a spherically symmetric relativistic
pre-shock wind with an arbitrary magnetization parameter
in the framework of ideal magnetohydrodynamics (MHD), while
  Romanova et al.\ (2005) postulated an equipartition between the
magnetic and particle energy densities and
 included  magnetic field reconnection into consideration.
 Because of computational  difficulties,
the numerical calculations  have been  so far limited to  short  distances
(a few shock stand-off radii)
 from the pulsar.
To understand the large-scale
properties of extended tails and facilitate their modeling
(including the intrinsic anisotropy of the wind and misalignment between
the pulsar velocity and spin),
observations of these objects are particularly important.

 Here we report
on
  X-ray observations of very extended pulsar tails
behind two middle-aged pulsars, J1509--5850 and J1740+1000.
 PSR J1509--5850 (hereafter J1509) was
  discovered  in the
Parkes Multibeam Pulsar Survey\footnote{
http://www.atnf.csiro.au/research/pulsar/pmsurv}
 (Kramer et al.\  2003).  The   pulsar is located
 in the Galactic plane at the dispersion measure (DM) distance
of
about 4 kpc
(see Table 1 for the pulsar properties).
To study the X-ray properties of the pulsar and look for
its PWN, we observed this field with \chan.
In addition to detecting the pulsar,  the observation
revealed
a spectacular
long tail, first
reported by Kargaltsev et al.\ (2006). It was
briefly described by Hui \& Becker (2007; hereafter HB07), who
also reported the discovery of
diffuse radio emission, possibly associated with the
tail. In this paper, we provide a detailed analysis of our \chan\
observation of the tail and discuss implications of our results.

Pulsar   J1740+1000 (hereafter J1740)
  has been discovered in an Arecibo survey (McLaughlin et al.\ 2000), and
 the follow-up radio observations were carried out by McLaughlin et al. (2002).
  The pulsar's
 properties are summarized in Table 1.
  At the DM distance of
1.4 kpc and  Galactic latitude of
 $20.3^\circ$, J1740 belongs to
 a very small
group of pulsars located  at  large distances from the
 Galactic
 plane,
 implying either a progenitor from the halo population or
a neutron star (NS) ejected from the Galactic disk with an exceptionally
high speed,
$\sim4000$ km s$^{-1}$.
J1740 was observed with \chan\ in 2001 (PI: Z.\ Arzoumanian).
  In a preliminary analysis of
these data we detected the pulsar, but the
exposure was too short to study
possible extended emission
and the pulsar's spectrum.
Therefore, we observed the
field around J1740 in  a much deeper imaging exposure
 with {\sl XMM-Newton}.
   In this paper
  we report the discovery
 of the PWN
associated with J1740 and focus on the  analysis
of the  PWN  properties,
 while the spectral and timing properties of the pulsar,
 including the detection of the thermal emission
 from the NS surface, will be presented in a separate publication
(Misanovic et al., in prep.).

Our X-ray observations of the J1509 and J1740 tails
 and the data analysis are described in \S2.
We estimate  the pulsar and flow velocities, the magnetic field
 strengths and
 energetics of the tails,
compare these tails with the other pulsar tails currently known, and  discuss
the implications of our findings in \S3. Our main results are
summarized in \S4.

\begin{figure*}[t]
 \centering
\includegraphics[width=6.5in,angle=0]{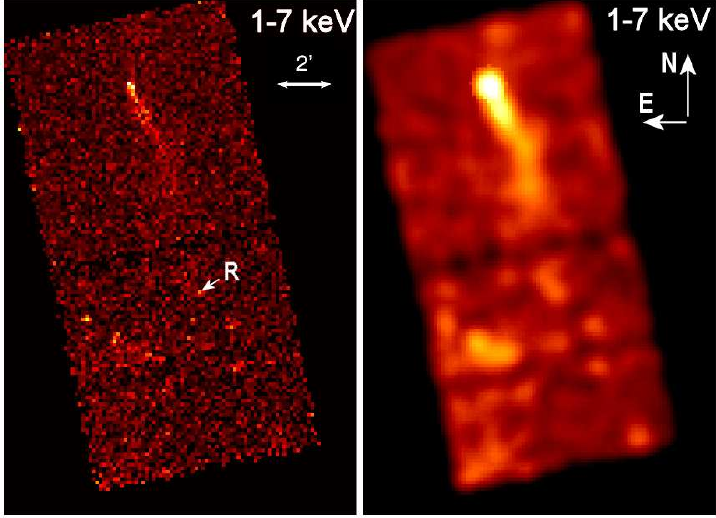}
 \caption{Large-scale ACIS view of  the J1509's X-ray tail.
   The images show
the S3 and S4 chips.
The 1--7 keV energy band is chosen to maximize the S/N  of the faint,
extended tail. Both images  are  binned by a factor
  of 16 (pixel size is $7.8''$). The image on the right is additionally smoothed with the $24''$ Gaussian kernel.
``R'' marks the counterpart of the pointlike source projected onto
the putative
radio tail (see \S3.2 and Fig.\ \ref{radio-xray-color}).
\label{tail-1509}}
\end{figure*}

\section{Observations and Data Analysis}

  J1509 was observed with the Advanced CCD Imaging Spectrometer (ACIS)
on board {\sl Chandra} on 2003 February 9 (ObsID 3513).
 The useful scientific exposure time was 39.6 ks. The
observation was carried out in Very Faint mode, and the pulsar was
imaged on the S3 chip,
$35''$ from the optical axis
(CHIPX=285, CHIPY=503).
 The other chips
 activated during this observation were S1, S2, S4, I2, and I3.
The detector was operated in Full Frame mode, which provides time
resolution of 3.24 s. For the analysis, we used the data
reprocessed on 2006 July 20  (ASCDS ver.\
7.6.8, CALDB ver.\
3.2.4).

J1740 was observed with the \chan\ ACIS on 2001  August 19 in
Timed Exposure (imaging) mode (ObsID 1989; useful exposure time 5.1 ks)
and Continuous Clocking mode (ObsID 2426; useful exposure time 16.5 ks).
Only the former observation
is useful for studying faint, extended features;
it was carried out in  Faint  mode with the
 pulsar
  imaged on the S3 chip,
$35''$ from the optical axis.
 The other activated chips were  S2, I0, I1, I2, and I3, and
  the time resolution was 3.24 s. For the analysis, we used the data
reprocessed on 2006 December 21 (ASCDS ver.\
7.6.9, CALDB ver.\
3.2.4).

 J1740 was also observed with the European Photon Imaging Camera
(EPIC) on board {\sl XMM-Newton} on 2006 September 28 and 30
(observations 0403570101 and 0403570201). The EPIC PN camera
was operated in  Small Window  mode with the reduced
 field-of-view (FOV)
 of $4'\times4'$ and   time-resolution of 6 ms,
to analyze pulsations and perform phase-resolved spectroscopy
of the pulsar.
 To obtain a large-scale image of the
PWN,
the EPIC MOS1 and MOS2 cameras were operated in
 Full Window Mode, which provides  a
larger FOV
at the expense of lower time resolution (2.6 s).
 The Thin filter was in front of the PN and MOS cameras during both
 observations.
 The total exposures were 40.4 and 25.6 ks for the first and  second
 observations, respectively.   However, because of
the reduced  efficiency of the PN detector in
  the Small Window mode, the effective PN exposures were 28.6 ks and
17.8 ks in the first
and second observations, respectively.
We did not find
significant flaring events in both observations
(the background
count rate varied by $\lesssim20\%$). Thus,
the  useful scientific exposures were
 66 ks for MOS1 and MOS2, and 46.4 ks for PN in the two observations combined.

The {\sl Chandra} data were reduced using the Chandra Interactive
Analysis of Observations (CIAO) software (ver.\ 3.4; CALDB ver.\ 3.4.0).
The data obtained by {\sl XMM-Newton} were reduced with the
  Scientific Analysis Software (SAS, ver.\  7.0).

\subsection{Images}

The X-ray images of
the J1509 and J1740 fields
  reveal point sources at the radio pulsar positions.
The best-fit centroid positions
 obtained from the
 {\sl Chandra} ACIS images,
  are ${\rm R.A.}=15^{\rm h}09^{\rm m} 27.163^{\rm s}$,
 ${\rm decl.}=-58^{\circ}50' 56.12''$,
and
 ${\rm R.A.}=17^{\rm h}40^{\rm m} 25.939^{\rm s}$,
${\rm decl.}=+10^{\circ}00' 05.78''$,
 for J1509 and J1740,
respectively. These positions differ from the corresponding radio
positions
 by $0.3''$ and $0.6''$,
 respectively. The differences are comparable
 to the
  uncertainty  of the absolute {\sl Chandra} astrometry\footnote{http://cxc.harvard.edu/cal/docs/cal\_present\_status.html\#abs\_spat\_pos}.

\begin{figure}
\begin{center}
\includegraphics[height=6.3cm,angle=0]{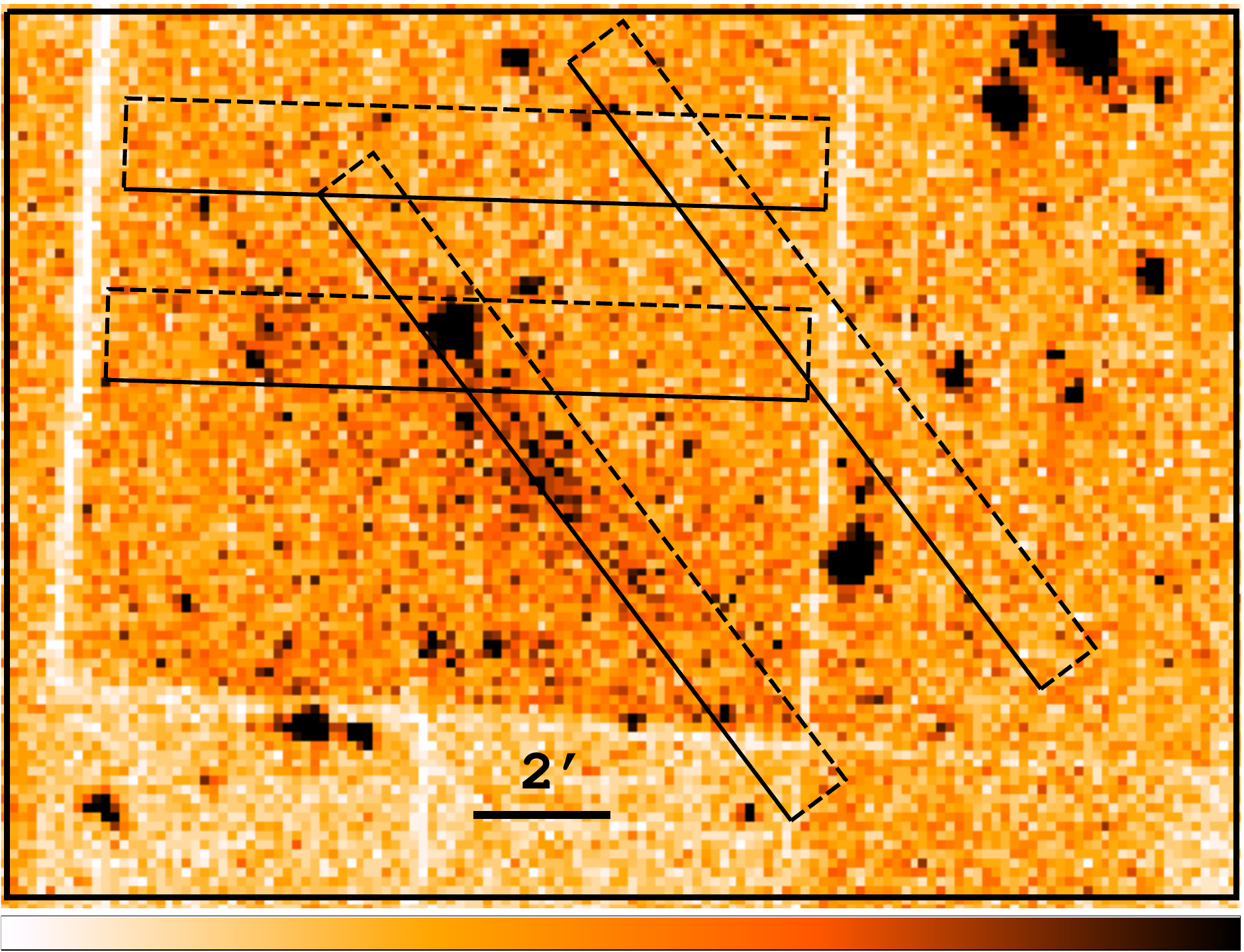}
\includegraphics[height=6.3cm,angle=0]{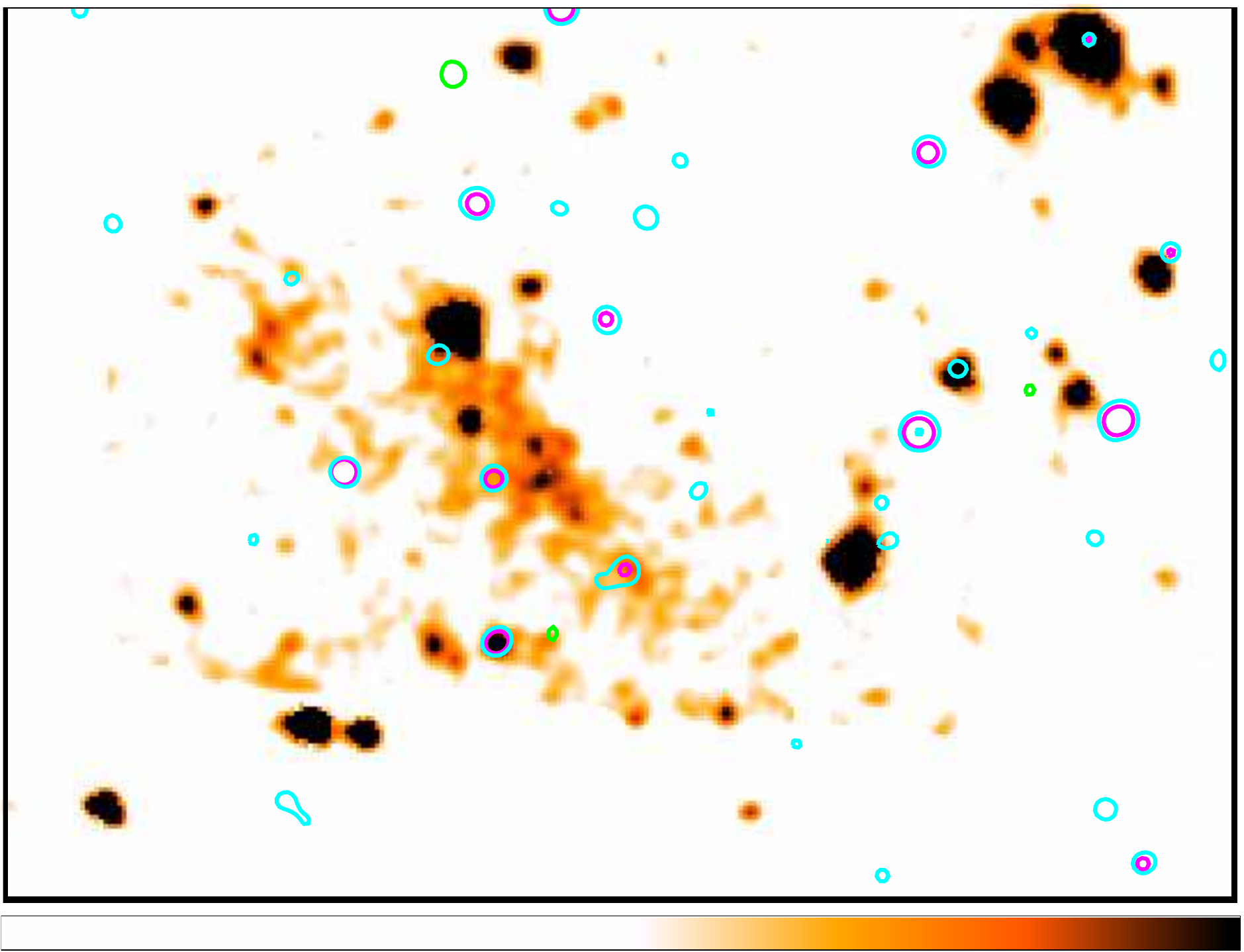}
\end{center}
\caption{
 Combined MOS1+MOS2  images ($18' \times 13'$;
0.3$-$10\,keV) of the field around
 J1740 from two {\sl XMM-Newton}
 observations (see text). The {\em top} panel shows a binned image
(pixel size $8''$) while the {\em bottom} image (pixel size $4''$) is
 adaptively smoothed with a Gaussian kernel.
 The top panel shows the regions from which the linear
brightness profiles
(shown in Fig.~\ref{1740-profile-tail}) were extracted.
 The overlaid contours in the bottom image show optical and radio
sources from the DSS2-blue
(shown in cyan), DSS2-red
(magenta), and NVSS
 1.4 GHz (green) images.
\label{tail-1740}}
\end{figure}

\begin{table*}[]
\caption[]{Count statistics for
 PWN regions}
\vspace{-0.5cm}
\begin{center}
\begin{tabular}{cccccccc}
\tableline\tableline Region & $A$   &
 $N_{\rm tot}$  &
 $N_{\rm bg}$   &  $N_{\rm src}$ & S/N & $\mathcal{S}$ \\
 \tableline
\tableline
&&&  J1509 &&& \\
\hline
          Head           &
         0.073 &  $118\pm11$     &  $18\pm1$   &  $100\pm11$ & 9.1  &
$34.6\pm 3.8$ \\
R1          &
      1.16   &  $542\pm23$     &  $249\pm 11$   &  $293\pm26$ & 11.3  & $6.38\pm0.57$ \\
R2          &
      1.26  &  $453\pm21$     & $280\pm12$    &   $173\pm24$ & 7.2  & $3.45\pm0.48$  \\
R3          &   2.24
              &  $691\pm26$     &  $530\pm16$   & $161\pm31$ & 5.2 & $1.80\pm0.35$ \\
         Entire tail    & 4.68
               &  $1686\pm41$      &  $1059\pm23$   &  $627\pm47$  & 13.3 &
 $3.38\pm 0.25$ \\

 \tableline
&&&  J1740 &&& \\
\hline
Bright portion  (MOS1+2)         & 2.10         &  $1658\pm41$      & $1026\pm32$    & $632\pm56$ & 11.4 & $6.48\pm0.57$  \\
Bright portion (PN)         & 2.10         &  $2445\pm49$      & $1791\pm42$    & $654\pm70$ & 9.2 & $6.70\pm0.72$  \\
      \tableline
\end{tabular}
\end{center}
\tablecomments{
Total ($N_{\rm tot}$),
scaled background ($N_{\rm bg}$), and source ($N_{\rm src}$) counts  are in the
0.5--8 keV band for the {\sl Chandra}
data,
 and 0.4--10 keV band for {\sl XMM-Newton} combined data, extracted from the
regions of area $A$ (in arcmin$^{2}$) shown in Figs.\ \ref{regions} and
\ref{tail-1740-mos-pn} for J1509 and J1740, respectively.
The mean surface brightness, $\mathcal{S}$, is in units of
 counts ks$^{-1}$  arcmin$^{-2}$ in the 0.5--8 keV and 0.4--10 keV bands
for J1509 and J1740, respectively.
\label{counts}}
\end{table*}

\begin{figure}
\begin{center}
\includegraphics[height=5.3cm,angle=0]{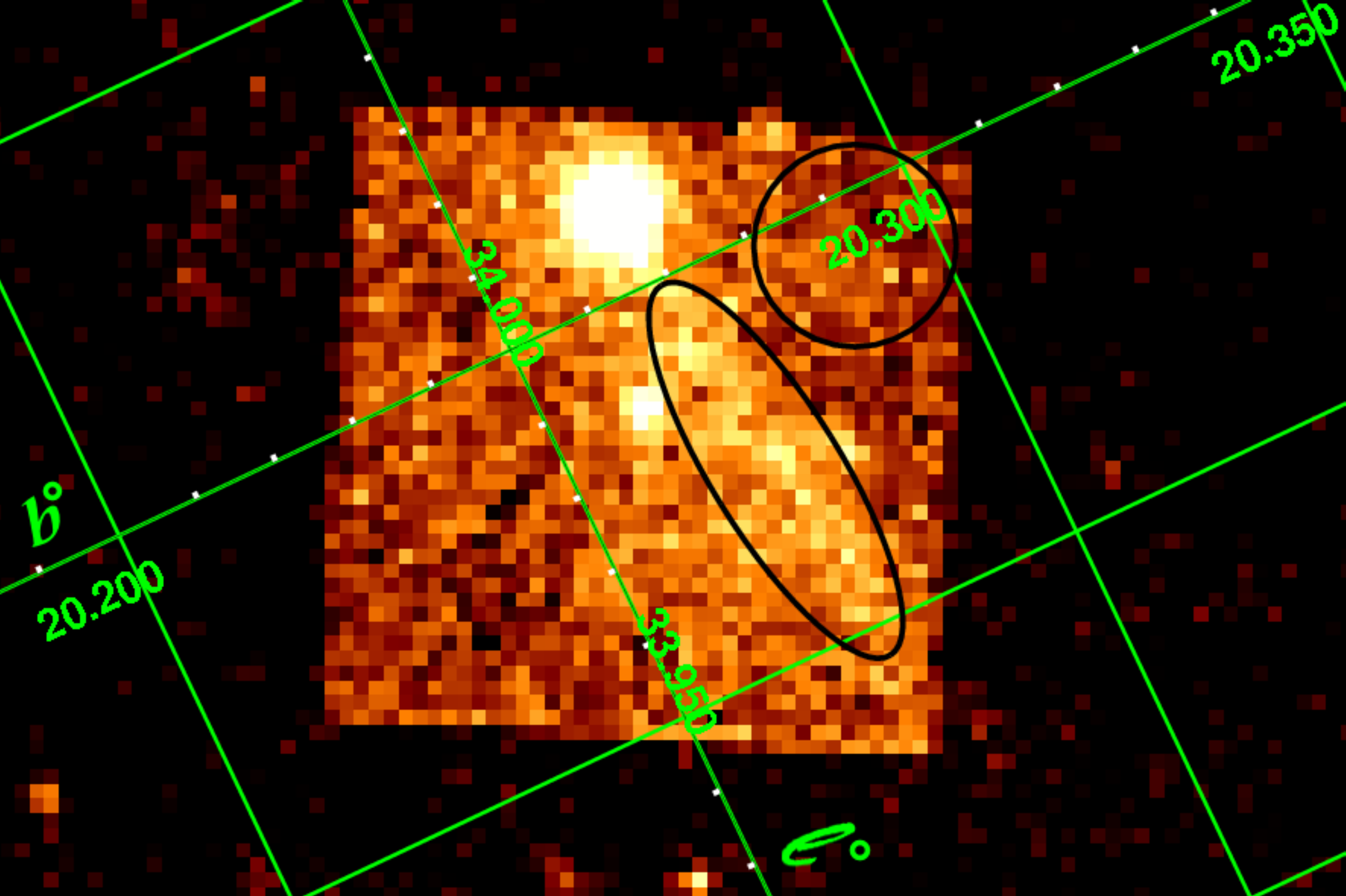}
\end{center}
\caption{
$9.4' \times 6.2'$
 binned image (pixel size $8''$) of the field around
 J1740 combining MOS1+MOS2 (Full Frame mode) and PN (Small Window mode) data
of two  {\sl XMM-Newton} observations, in the 0.3$-$10\,keV band (see text).
  The regions used for  spectral extraction are
 marked by an ellipse (the source region) and a circle (the background). Galactic coordinate grid is shown.
 \label{tail-1740-mos-pn}}
\end{figure}

  The most striking features
   exposed  in the binned {\sl Chandra} and {\sl XMM-Newton} images
(see Figs.\ \ref{tail-1509}, \ref{tail-1740},
 and \ref{tail-1740-mos-pn}) are
long linear structures  (we will call them ``tails'' hereafter)
 attached
to  J1509 and J1740.
  The tails have similar orientations on the sky, with the position angles of
$\approx 203^{\circ}$  for  J1509, and $\approx 215^{\circ}$  for J1740
(counted east of north). Assuming that the tails are caused by the pulsar
motion, the corresponding
position angles of the pulsars' proper motion can be estimated
as $\approx 23^\circ$ and $\approx35^\circ$.

\begin{figure}[t]
 \centering
\includegraphics[width=3.2in,angle=0]{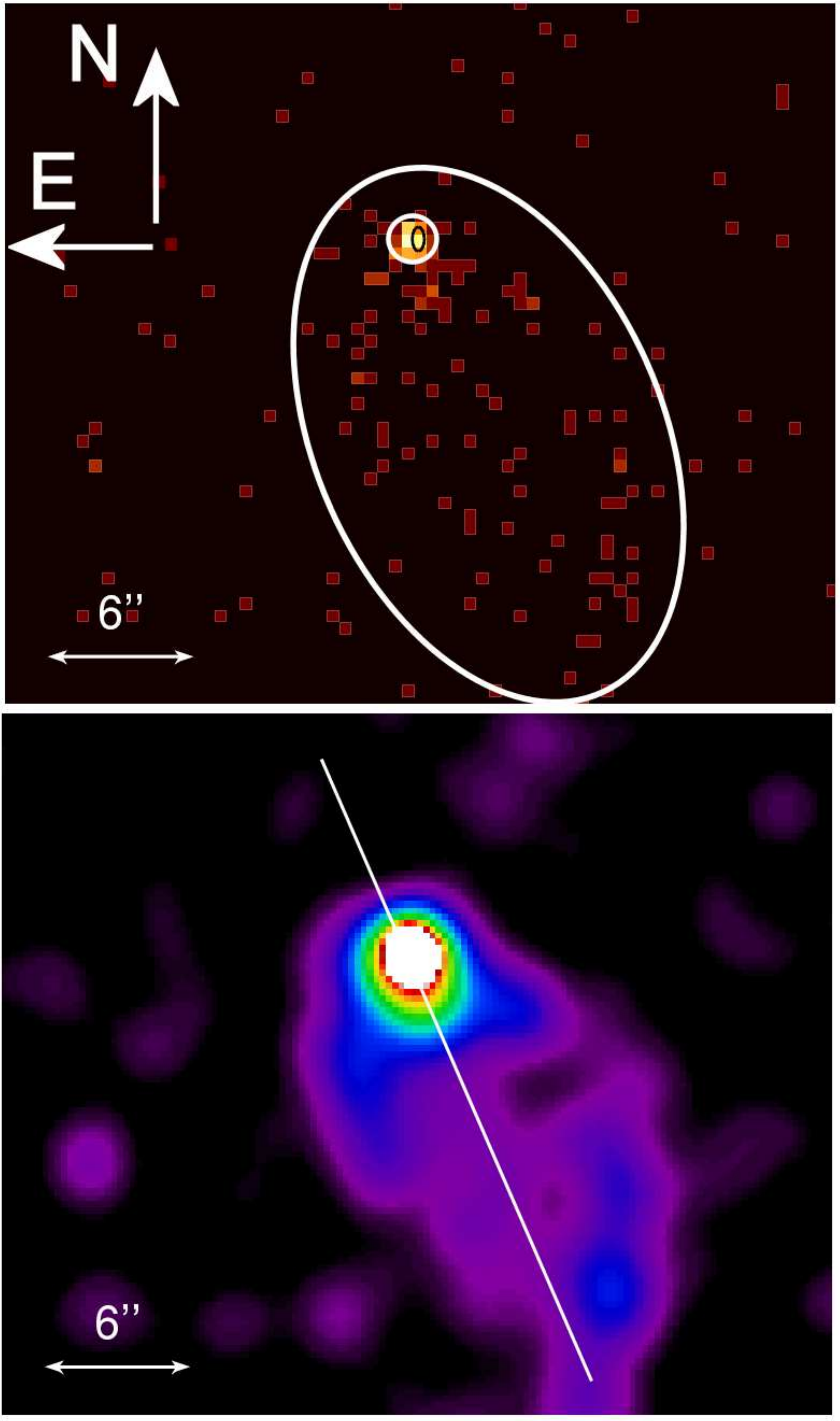}
 \caption{  {\em Top:}  ACIS-S3 image
 of J1509 and its vicinity (1--7 keV; pixel size $0\farcs49$). Also
shown are the extraction
regions used to measure the pulsar spectrum ($r=0.9''$ circle)
 and the spectrum of the PWN head
($14''\times 24''$ ellipse). The radio pulsar position
 is shown by the small black ellipse, whose size
shows the position uncertainty.
 {\em Bottom:} Adaptively smoothed sub-pixel
resolution image (pixel size $0.25''$)
 of the same region obtained by removing the
pipeline pixel randomization and subsequently applying the sub-pixel
resolution tool (based on analyzing the charge distribution produced
by an X-ray event; Tsunemi et al.\ 2001; Mori et al.\ 2001).  The line
shows the approximate symmetry axis of the PWN, which can be crudely
identified with the direction of pulsar's proper motion.}
\label{tail-vicinity-1509}
\end{figure}

\subsubsection{J1509 tail}
\begin{figure*}
 \centering
\includegraphics[width=6.7in,angle=0]{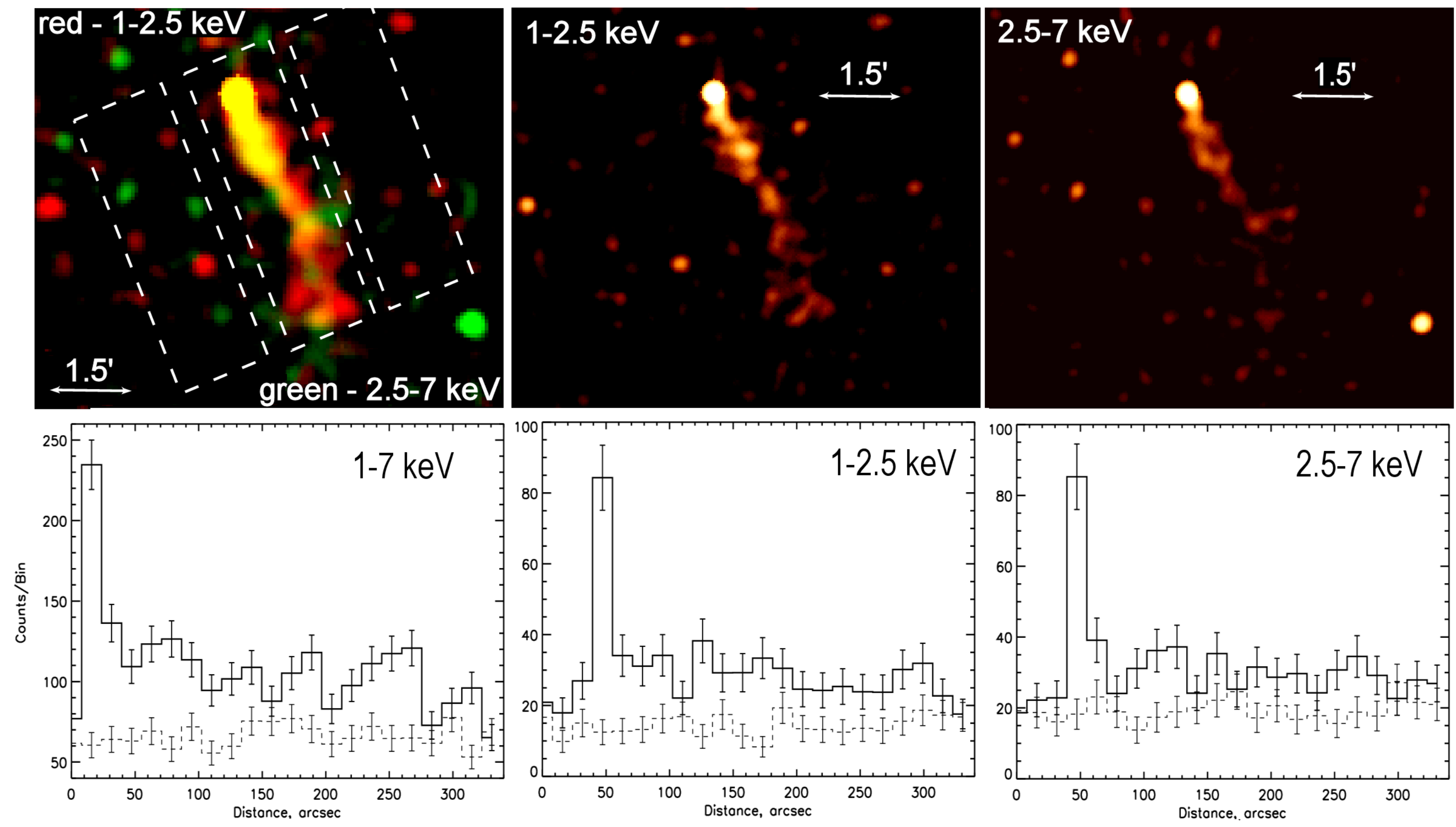}
\caption{ACIS  images ({\em top}) of the J1509 tail
(binned to the pixel size of $4''$ and smoothed with a $20''$
Gaussian kernel) in the total (1--7 keV; {\em left}),
soft (1--2.5 keV; {\em middle}), and hard (2.5--7 keV; {\em right}) bands.
The total band image is color coded, with the
soft emission shown in red and hard emission shown in green.
The {\em bottom} panels show the tail profiles in each energy band.
The linear brightness profiles
  are obtained by extracting the source and background counts from the $1.6'\times6'$ rectangular  regions  shown in the {\em top left} panel. The original unsmoothed
  images  were used for the extraction.
The profile extracted from the source region is plotted
by the solid line while the
  averaged background profile extracted from the adjacent  background
regions is shown by the dashed line.
 The bin size is $15''$ in the linear profile plots.}
\label{1509-energy-resolved}
\end{figure*}

\begin{figure}
 \centering
\includegraphics[width=3.2in,angle=0]{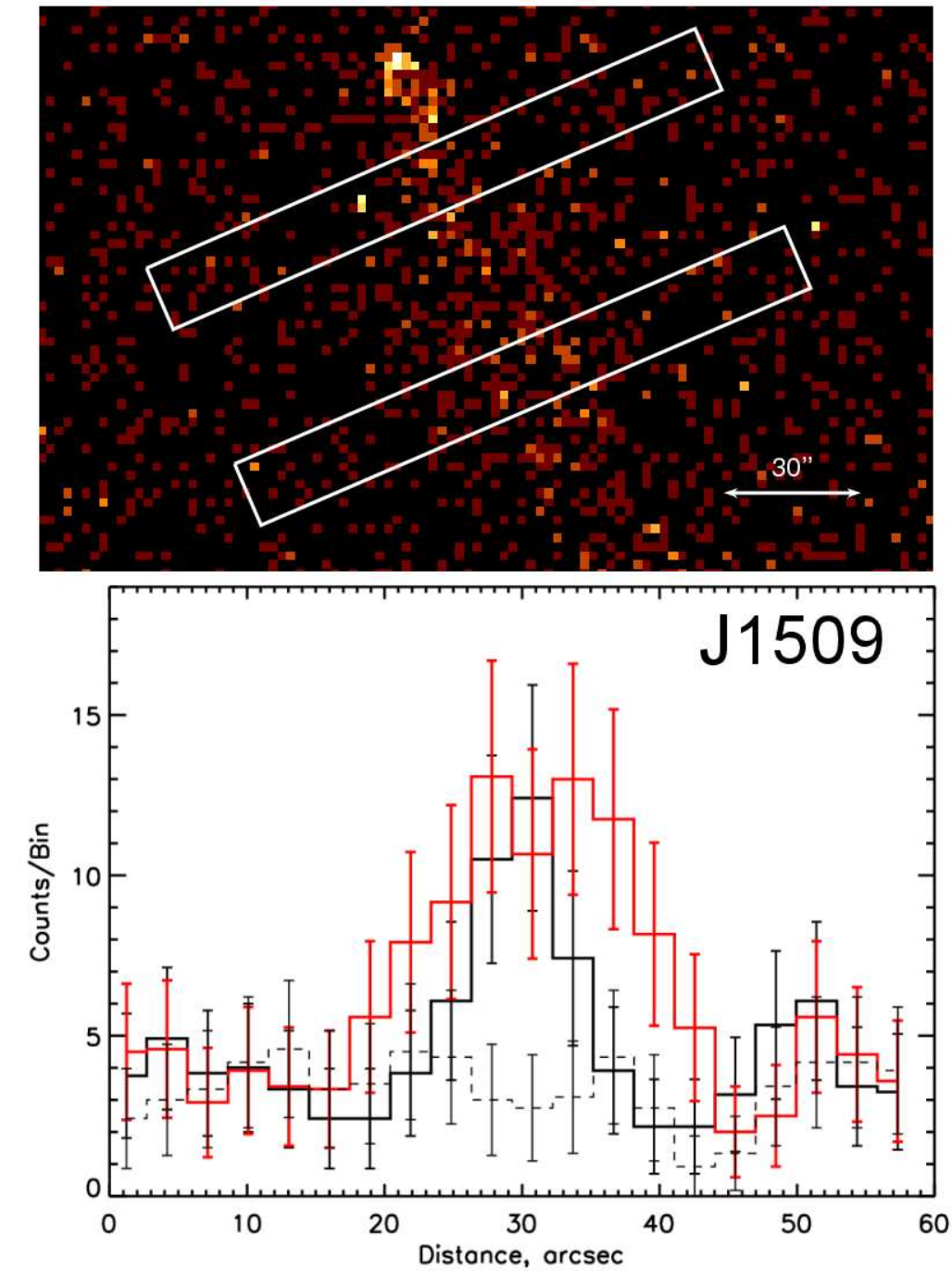}
\caption{{\em Top:}  ACIS-S3 image
(1--7 keV)  showing the brightest part of the J1509 tail.
 {\em Bottom:} Linear profiles across  the tail obtained by extracting
the source counts from the two rectangular ($132''\times15''$) regions
  shown in the top image. The original unsmoothed
1--7 keV  image
was used for the extraction. The profiles extracted from the regions closer to
and farther from the pulsar  are plotted by the black and red histograms,
respectively.
The dashed black histogram shows the
  averaged background profile extracted from the two background regions
on both sides of the tail.
\label{1509-profile2}  }
\end{figure}

\begin{figure}
 \centering
\includegraphics[width=2.5in,angle=0]{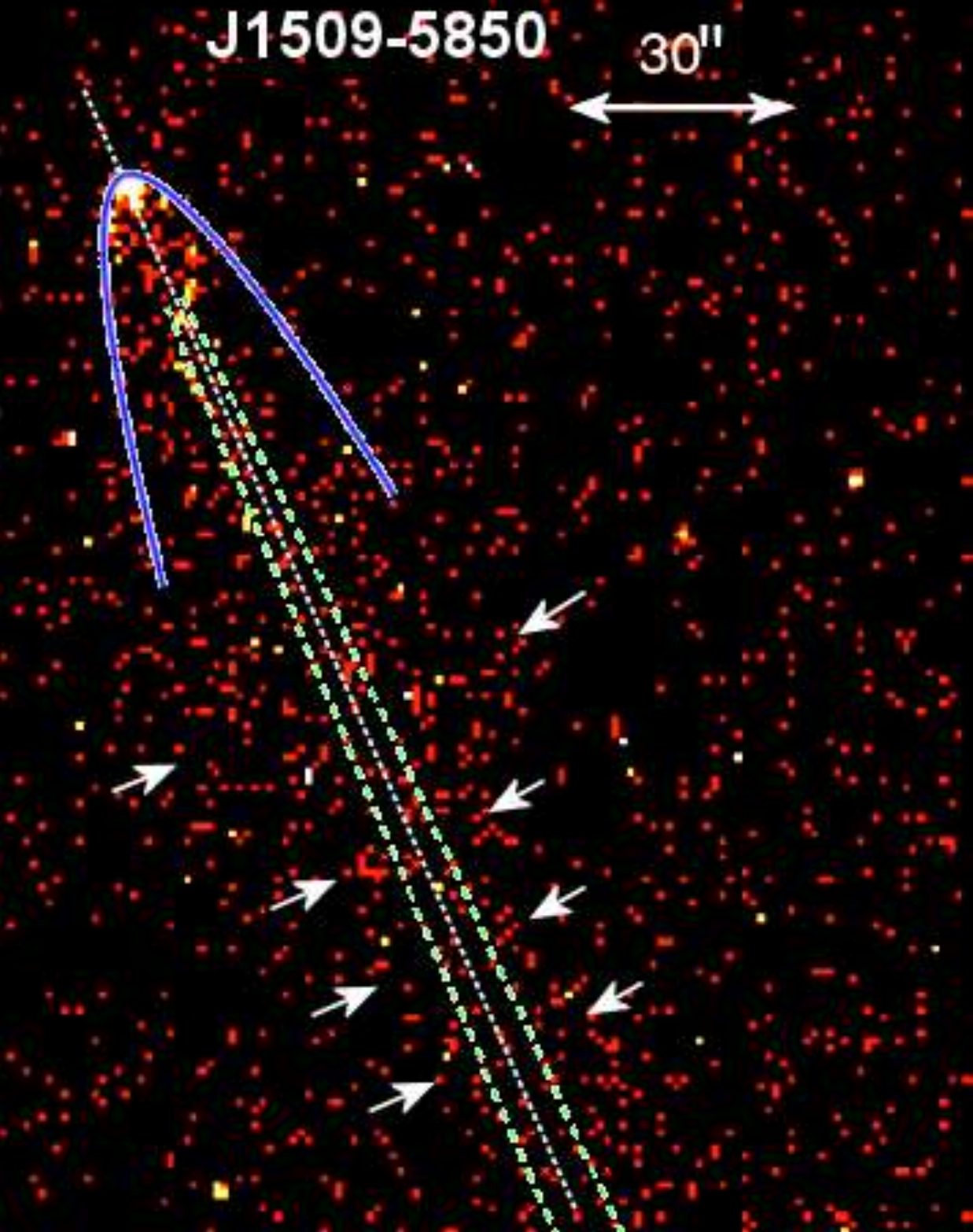}\\
\vspace{3mm}
\includegraphics[width=2.5in,angle=0]{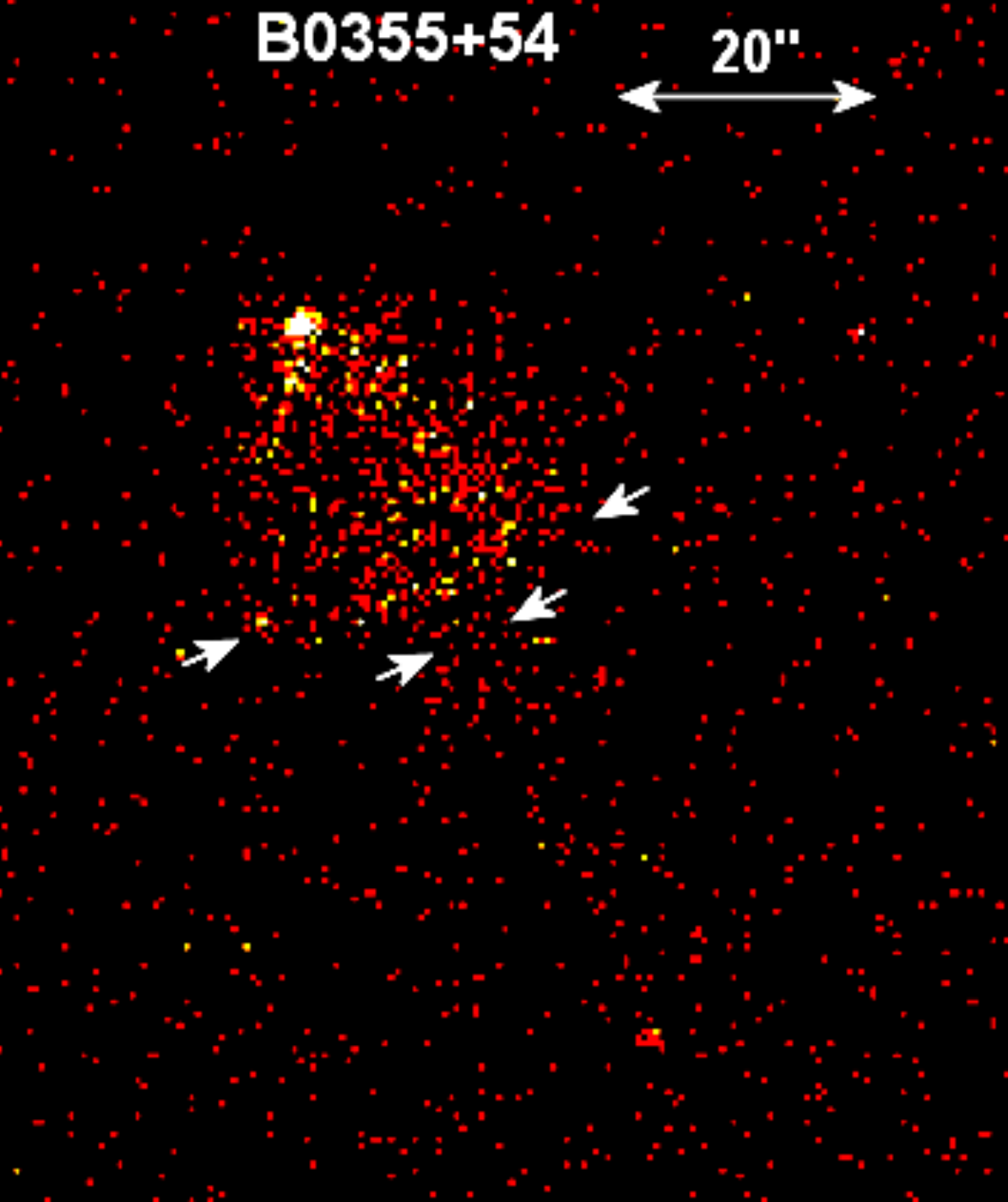}
\caption{
 {\em Top:} Images of the J1509 tail (in the 1--7 keV band,
binned to the pixel size of $2''$).
 The parabola, $x=2.45'' [(z+0.5'')/1'']^{1/2}$, where $z$ is the distance
from the pulsar
along the tail, shows an approximate boundary of the PWN head within
$\sim 15''$ from the pulsar, but it does not fit the tail's shape
at larger distances.
The part of the other parabola, $x=0.44'' (z/1'')^{1/2}$ (shown by the dashed line),  corresponds
to eqs.\ (32--34) in Romanova et al.\ (2005),
 for $v_{mz}=10,000$ km s$^{-1}$, $v_{\rm psr}=300$  km s$^{-1}$
 and $r_{0}=0.5''$ (see also \S3.2).
The white arrows
show an approximate
observed width of the tail (see text).
 {\em Bottom:}
The Mushroom PWN powered by PSR B0355+54.
Notice the sudden narrowing
(transition from the ``cap'' to the ``stem''),
similar to that seen in the J1509 tail.
\label{wilkin}  }
\end{figure}

The
 ACIS image
 of J1509  reveals the
compact PWN  ($\approx10''\times30'' $) in the
vicinity of the pulsar (see Fig.\ \ref{tail-vicinity-1509}).
 Because of its elongated, elliptical shape, we  will call  it
the  ``head'' hereafter.
 Being the brightest close to the
pulsar, the  diffuse emission gradually broadens and becomes fainter
with increasing distance from the pulsar,
 forming an extended tail. The tail emission is
 firmly detected  up to
 $\approx 5.6'$ (or $l_\perp\approx6.5 d_4$ pc, where $d_4=d/4\,{\rm kpc}$)
(see Fig.\ \ref{tail-1509}) and possibly even
 farther, at a lower significance.
 This is
 demonstrated by the linear
surface brightness profiles extracted along and across the
tail (see Figs.\ \ref{1509-energy-resolved} and \ref{1509-profile2}).
Although the collected number of counts from the
 tail is not very large  (see Table~\ref{counts}), the
low ACIS background still allows us to detect interesting
 morphological changes along the tail.
For instance,
beyond $\sim 15''$
from the pulsar,
the distribution of counts  in the tail becomes
   asymmetric  with respect to the inferred pulsar proper motion direction,
shown by the dashed straight line in Figure \ref{wilkin}
(e.g.,   there
are more counts southeast of the line than northwest of the line).
Being $\sim40''$ wide at the $\approx1.3'$ distance from the pulsar,
   the tail slightly brightens
and then suddenly   narrows by  about  a factor of two
(as shown by the small arrows in the top panel of Fig.~\ref{wilkin}).
 Interestingly, very similar behavior is observed in the
``Mushroom PWN'' (KP08)
associated with
 another middle-aged pulsar,  B0355+50 (see
Fig.\ \ref{wilkin}, {\em bottom},
and the images in McGowan et al.\ 2006).
 Local tail brightenings are also seen  at larger distances from
the J1509 pulsar
 (Figs.\ \ref{1509-energy-resolved} and \ref{1509-profile2});
 however,
the decreasing
surface brightness of the tail
and reduced  off-axis resolution do not allow one
to
check whether or not these brightenings
 are accompanied by significant morphological changes.

\subsubsection{J1740 tail}
\begin{figure}
\begin{center}
 \includegraphics[height=6.3cm,angle=0]{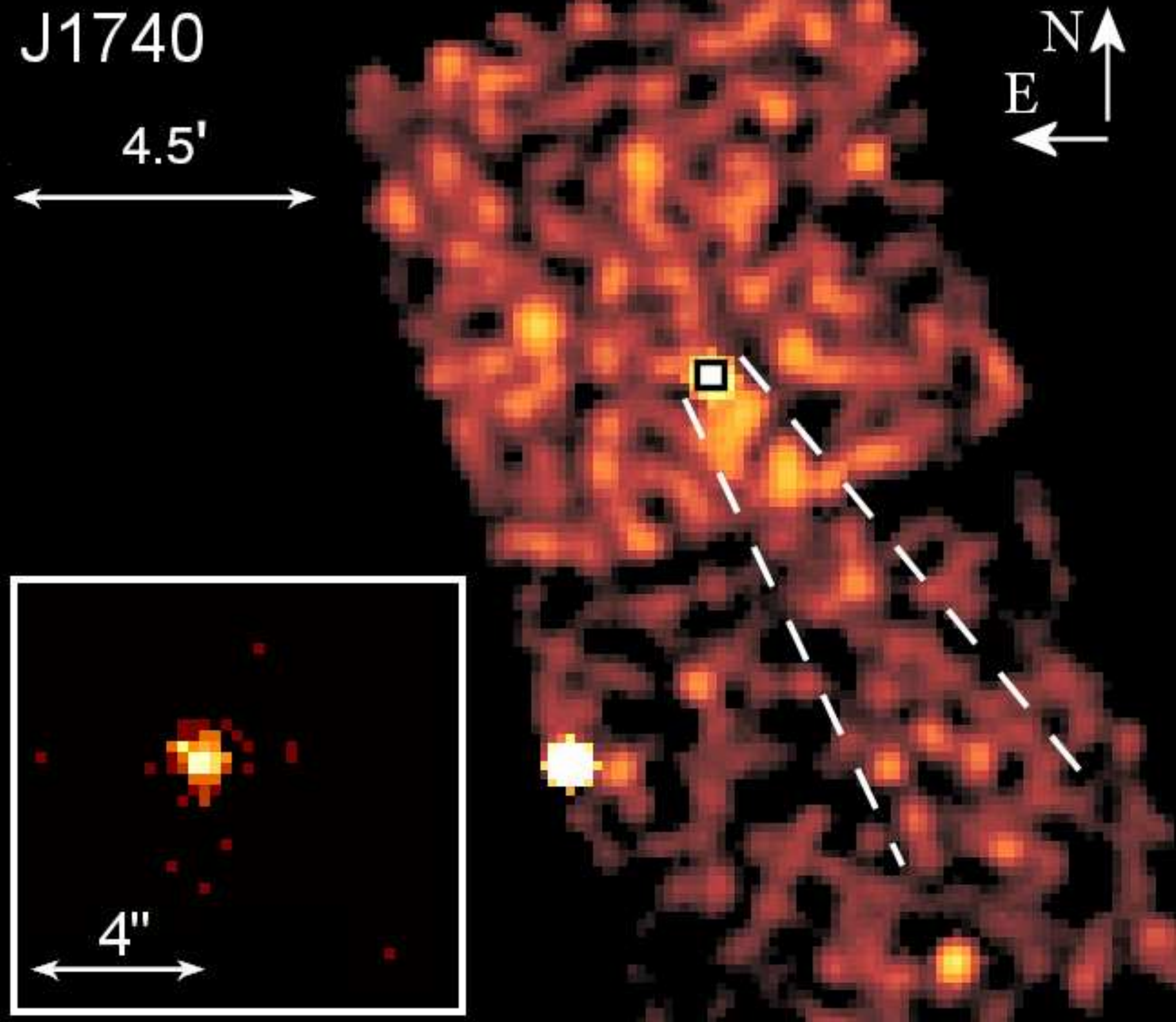}
\end{center}
\caption{ ACIS images of the J1740 field. The main panel shows
the binned (pixel size $7.9''$) ACIS image
in the 0.5--6 keV band
 smoothed with a $24''$ Gaussian.  Faint emission from the
extended tail is
 visible within the region between the dashed lines.
  The inset
shows the immediate vicinity
of J1740 in the same energy band.
The image resolution is maximized by applying a sub-pixel
resolution tool
(Tsunemi et al.\ 2001; Mori et al.\ 2001).
The pixel size in the inset  image is $0.25''$.
\label{1740-ACIS}}
\end{figure}

\begin{figure}
\begin{center}
\includegraphics[height=10.0cm,angle=0]{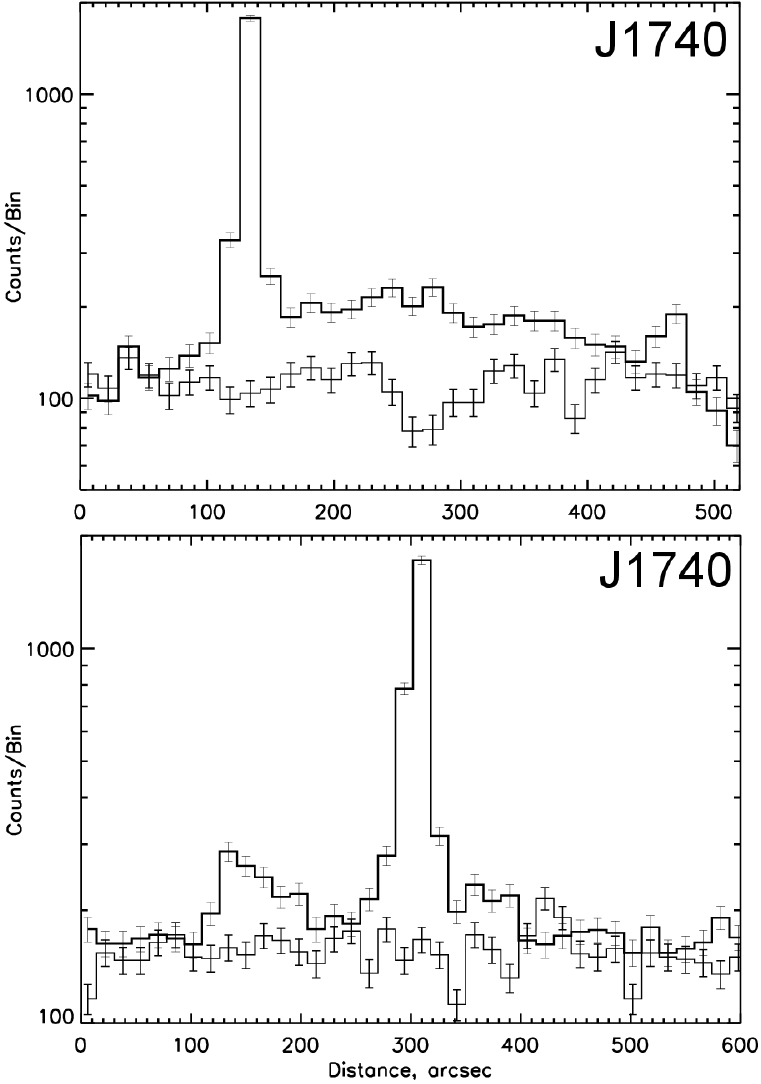}
\end{center}
\caption{
{\sl  Top:} Linear profiles of the J1740 tail and the background
obtained by extracting counts
from the rectangular regions along the tail and  a parallel source-free
region located  north-east from the tail
(the two $1' \times 11.5'$ rectangular regions are shown in
Fig.\ \ref{tail-1740}).
{\sl Bottom:} Linear profiles extracted along the east-west direction  from
 the  two $1.3' \times 10.3'$  rectangular regions shown in
Fig.\ \ref{tail-1740}. The unsmoothed MOS1+MOS2 image in the 0.4--10\,keV band,
 combining
the two {\sl XMM-Newton} observations, was used for the extraction.
\label{1740-profile-tail}}
\end{figure}

Even in the short, 5 ks,
ACIS imaging exposure of J1740
we can see traces of
faint extended emission
southwest of the pulsar
in the heavily
 binned  ACIS-S image
(Fig.\ \ref{1740-ACIS}),
 and  the unbinned image of the immediate vicinity of the pulsar
 shows an excess of counts in the same
direction
and perhaps some emission in front of the pulsar
 (see the inset in Fig.\ \ref{1740-ACIS}).
  The
  long ($\sim5.5'$, or $l_\perp \sim2d_{1.4}$\,pc, where $d_{1.4}=d/1.4\,{\rm kpc}$)
tail is much better   seen in the EPIC
images
(Figs.\ \ref{tail-1740} and \ref{tail-1740-mos-pn}).
 Similar to J1509,
 the linear profile of the J1740 tail (see Fig.\ \ref{1740-profile-tail})
hints at  surface brightness
 fluctuations along the tail;
 however, detailed investigation is not possible
because of the high EPIC background and poor angular resolution
 of {\sl XMM-Newton}.
 In addition to the long tail, the smoothed MOS1+MOS2
image of the J1740 field (Fig.\ \ref{tail-1740})
and the linear profile in the east-west
direction (Fig.\ \ref{1740-profile-tail}) show some diffuse emission
$\approx 2.5'$ east of the pulsar, whose connection to the pulsar
remains unclear. Since the {\sl XMM-Newton} resolution
is much more coarse
than that of {\sl Chandra},
emission from a number of clustered point sources may sometimes
mimic the
diffuse emission in the
EPIC images.
 Therefore, we searched for point sources at radio and optical wavelengths.
 The bottom panel of
Figure \ref{tail-1740} shows the contours of optical and radio
sources detected in the DSS2-blue, DSS2-red and NVSS (1.4 GHz)
images of the
region around J1740. There are a few
optical
 sources
that coincide with the X-ray pointlike sources in the
 EPIC images. However, the majority of optical sources
 in the field
are faint and have no
X-ray counterparts.
Furthermore, the distribution of these sources
shows that
they cannot be responsible for the
emission detected in the tail region, nor
in the region $\approx 2.5'$  east of the pulsar.
 Yet, until a high-resolution, deep ACIS image is obtained, we cannot
exclude the possibility
 that some of the J1740 tail emission comes from faint point sources
lacking obvious optical counterparts.

As the orientation of the J1740 tail on the sky indicates the
proper motion direction, it provides useful information
about the pulsar's birthplace. Being located high above the Galactic plane,
at $z=0.48\, d_{1.4}$ kpc,
 the pulsar
might be a very high-speed NS if it were born in the plane
(e.g., $v\gtrsim
4200\, d_{1.4}\, (114\,{\rm kyr}/t)$ km s$^{-1}$,
where $t$ is the pulsar
age, if it were born in the midplane of the Galaxy).
However,
the X-ray image of the tail,  shown together with the Galactic coordinate
grid  in Figure \ref{tail-1740-mos-pn}, suggests  that the pulsar is moving
 at a small angle, $\simeq7^{\circ}$,
{\em toward the
plane}, which
  means that
it was born well out of the Galactic plane, likely from a halo-star
progenitor.

\subsection{PWN spectra}

\subsubsection{J1509}

\begin{figure}
 \centering
\includegraphics[width=3.2in,angle=0]{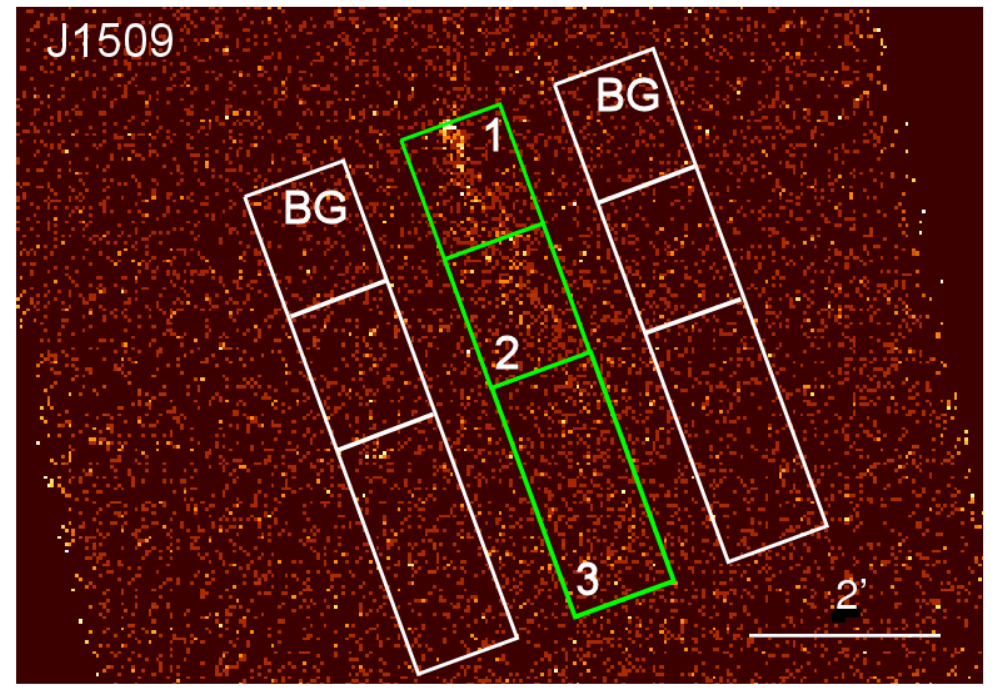}
\caption{ Regions used to extract the spectra of the J1509 tail  (see \S2.2.1).
 \label{regions} }
\end{figure}

  We extracted the  spectrum of the
J1509 PWN from the three numbered rectangular
 regions shown in Figure \ref{regions}.
Region 1 (R1 hereafter) includes the PWN head, but not the pulsar
(we excluded counts from the $0.9''$ radius circle shown in
Fig.\ 4).
  We also
  extracted the spectrum from the combined region 1+2+3
  (although it includes the PWN head,
 the head contribution is small compared to that of the tail,
so that we will refer to it
   as the ``entire tail'' region).
For each of the regions, the background was extracted
 from the
 two corresponding rectangular regions located
  on both sides of the source region (see Fig.\ \ref{regions}).
 The measured source and background count numbers and areas
  are given in Table \ref{counts}.
The  observed flux in the entire tail
is $F_{\rm tail} =(2.0\pm0.2)
\times 10^{-13}$ ergs s$^{-1}$ cm$^{-2}$ in
the 0.5--8 keV band. In addition, we extracted
the head spectrum
 from the elliptical region shown in
Figure \ref{tail-vicinity-1509} ({\em middle panel}),
 with the absorbed flux  $F_{\rm head} = (4.9\pm0.5)
\times 10^{-14}$ ergs s$^{-1}$ cm$^{-2}$ in the 0.5--8 keV band.

\begin{table}[]
\caption[]{PL fits to the J1509 and J1740 PWN spectra} \vspace{-0.5cm}
\begin{center}
\begin{tabular}{cccccccc}
\tableline\tableline Model & $n_{\rm H,22}$\tablenotemark{a}   &
$\mathcal{N}_{-5}$\tablenotemark{b}  &
$\Gamma$ & $\chi^2$/dof   & $L_{\rm X,32}$\tablenotemark{c} & $I_{-17}$\tablenotemark{d}  & $\mathcal{B}_{-8}$\tablenotemark{e}  \\
\tableline
&&&J1509&&&&\\
 \hline
   Head           &       $[1.3]$     &
$0.7\pm0.2$       &
$1.27\pm0.25$          &  $7.56/9$  & $1.15\pm0.32$  &
23 & 2.7 \\
  Head           &       $[2.1]$     &
$1.4^{-0.17}_{+0.19}$       &
$1.83\pm0.30$          &  $8.37/9$  & $1.33\pm0.41 $  &
27 & 5.3 \\
R1          &       $[2.1]$     &
$5.7_{-1.0}^{+1.2}$       &
$2.12^{-0.22}_{+0.19}$          &  $3.60/8$  & $4.4\pm0.4 $  & 5.5 & 1.4\\
R2         &       $[2.1]$     &
$4.3_{-1.2}^{+1.7}$       &
$2.36\pm0.40$          &  $1.68/4$  & $2.9\pm 0.6 $  & 3.7 &
0.95 \\
R3         &       $[2.1]$     &
$4.7\pm1.6$       &
$2.37^{-0.43}_{+0.35}$          &  $2.16/4$  & $3.2\pm 0.7 $  & 2.0 & 0.58 \\
  Entire tail        &       $[2.1]$                     &
$16.5\pm2.4$ &       $2.36_{-0.17}^{+0.15}$   &  $9.1/14$  & $11.2\pm1.0 $ & 3.5
 & 1.0 \\
      \tableline
&&&J1740&&&&\\
\hline
  Bright tail       &     $0.05^{+0.11}_{-0.05}$   & $0.73^{+0.25}_{-0.16}$ &
 $1.37^{+0.20}_{-0.25}$ &
50.9/38 & $0.16\pm0.04$ & 0.91 & 0.1 \\
  Bright tail     &     [0.1]   & $0.80\pm0.11$  &
 $1.48\pm0.19$ &
 51.5/39 & $0.15\pm0.04$ & 0.88 & 0.1 \\
\tableline
\end{tabular}
\end{center}
\tablecomments{
 The uncertainties are given at the 68\%
confidence level for one interesting parameter.
}
\tablenotetext{a}{The fits for J1509 are for fixed $n_{\rm H,22}\equiv
n_{\rm H}/10^{22}$ cm$^{-2}$. For J1740, the fits are for both
 free and fixed $n_{\rm H,22}$ (fixed values are in brackets).}
\tablenotetext{b}{Spectral flux in units of $10^{-5}$
 photons cm$^{-2}$ s$^{-1}$ keV$^{-1}$ at 1 keV.
}
\tablenotetext{c}{ Isotropic luminosity  in units of $10^{32}$ ergs s$^{-1}$ in
the 0.5--8 keV band for J1509 and
0.4--10 keV band  for J1740.
 }
\tablenotetext{d}{Mean unabsorbed
 intensity  in
units of $10^{-17}$ ergs s$^{-1}$ cm$^{-2}$ arcsec$^{-2}$ in the 0.5--8
keV band for the J1509 tail, and  in the 0.4--10 keV band for
the J1740 tail.}
\tablenotetext{e}{
 $\mathcal{B}_{-8}=\mathcal{B}/(10^{-8}\, {\rm photons}\, {\rm s}^{-1}\,
{\rm cm}^{-2}\, {\rm arcsec}^{-2})$; $\mathcal{B}= \mathcal{N}/A$
is the average
spectral surface brightness at $E=1$ keV.}
\label{spectral-1509}
\end{table}

\begin{figure}
 \centering
\includegraphics[width=3.1in,angle=180]{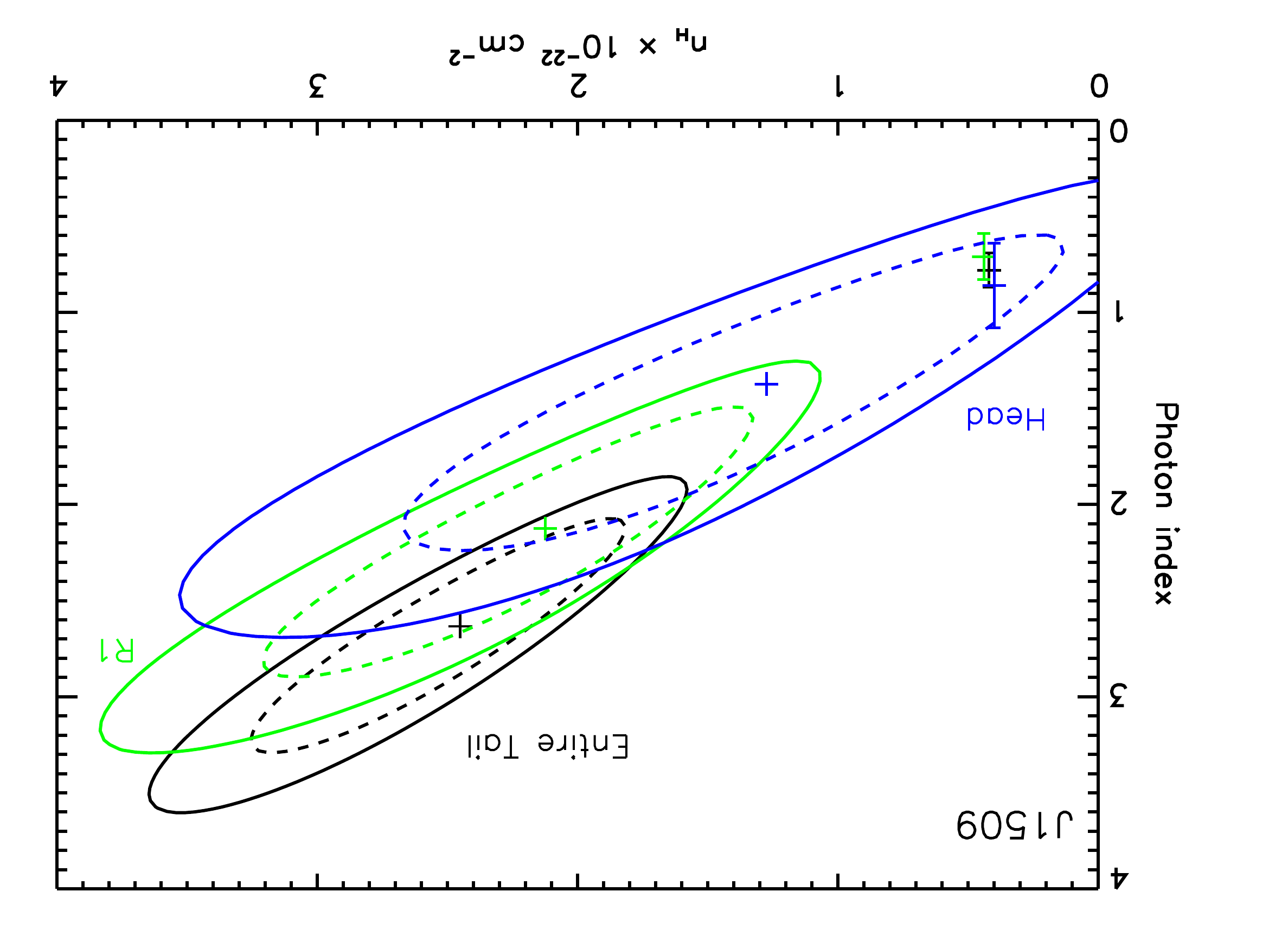}
 \caption{  Confidence contours (68\% and 90\%) in the $n_{\rm
H}$--$\Gamma$ plane for the PL fit to the J1509  head  ({\em blue}),
region R1
({\em green}),
 and the entire tail ({\em black}) spectra. The contours
 are obtained with the PL normalization fitted at each point
of the grid. The
$\Gamma$ values obtained from the fits  with
fixed $n_{\rm H,22}=0.42$ (inferred from the DM value assuming 10\% ISM
ionization) are
shown by the error bars.
}
\label{1509-contours1}
\end{figure}

First, we fit the R1 spectrum
with the absorbed
power-law (PL) model, allowing the hydrogen column density, $n_{\rm
H}$, to vary. The fit gives $n_{\rm H,22}\equiv n_{\rm H}/10^{22}\,
{\rm cm}^{-2} =2.1_{-0.5}^{+0.7}$ (the uncertainties here and below correspond to the 68\%
confidence level for a single interesting parameter, i.e.,
$\chi^2-\chi_{\rm min}^2 =1$). Fitting the same model to the R2 and R3 spectra
results in larger uncertainties, while the best-fit
$n_{\rm H}$ values are similar to that obtained for R1.
  The PL fit to the spectrum of the entire tail gives a slightly
larger column density, $n_{\rm H,22}=2.5\pm0.5$, while
 the fit to the head spectrum gives
a somewhat lower
$n_{\rm H,22}=1.3\pm0.5$,
  which is still consistent with the  above  values
(see Fig.\ \ref{1509-contours1})\footnote{
   These
values are  larger than $n_{\rm H,22}=0.82^{+0.93}_{-0.37}$
 reported by HB07 from the same ACIS data.
   From the description in HB07,
 it appears that the lower $n_{\rm H}$ was a result of
overcorrecting for the ACIS filter contamination.
To correct for the contamination,
HB07 used the XSPEC model ACISABS
 despite the fact that
this correction is already included in the ARF files generated by CIAO
(since the release of
  CALDB  2.26 on 2004 February 2). }.
   The total Galactic HI column in that direction is
$n_{\rm HI, 22}\approx1.7$
(Dickey \& Lockman 1990); however, this does
not contradict
 the
larger $n_{\rm H}$ values for the J1509 PWN
because the $n_{\rm H}$ deduced from an X-ray spectrum under the
assumption of standard element abundances generally exceeds the
$n_{\rm HI}$ measured from 21 cm observations by a factor of 1.5--3
(e.g., Baumgartner \& Mushotzky 2006).
Finally, the value $n_{\rm H,22}=0.42$,
estimated from the  pulsar's
dispersion measure
under the standard assumption of 10\% ionization of the interstellar
medium (ISM),
is noticeably lower than the above values, a
which possibly means that
 average ISM ionization
 is lower than 10\% in the direction to J1509.
   Thus, given different kinds of
uncertainties, we can only conclude that the true $n_{\rm H,22}$ is
somewhere between 0.4 and 3. Throughout most of the analysis below we choose
to fix  $n_{\rm H,22}$ at the value
 obtained  from the fit to the R1 spectrum since
it has the highest S/N among the other PWN elements
(see Table \ref{counts})\footnote{ Although
 the spectrum extracted from the  entire tail region has
even higher S/N and formally fits with a PL model,
  the actual spectrum of the entire tail
   can  deviate from a simple
 PL because of synchrotron cooling.
(Evidence of synchrotron cooling is seen in
the energy-resolved images shown in Fig.\ \ref{1509-energy-resolved}.)}.

\begin{figure}
 \centering
\includegraphics[width=3.1in,angle=180]{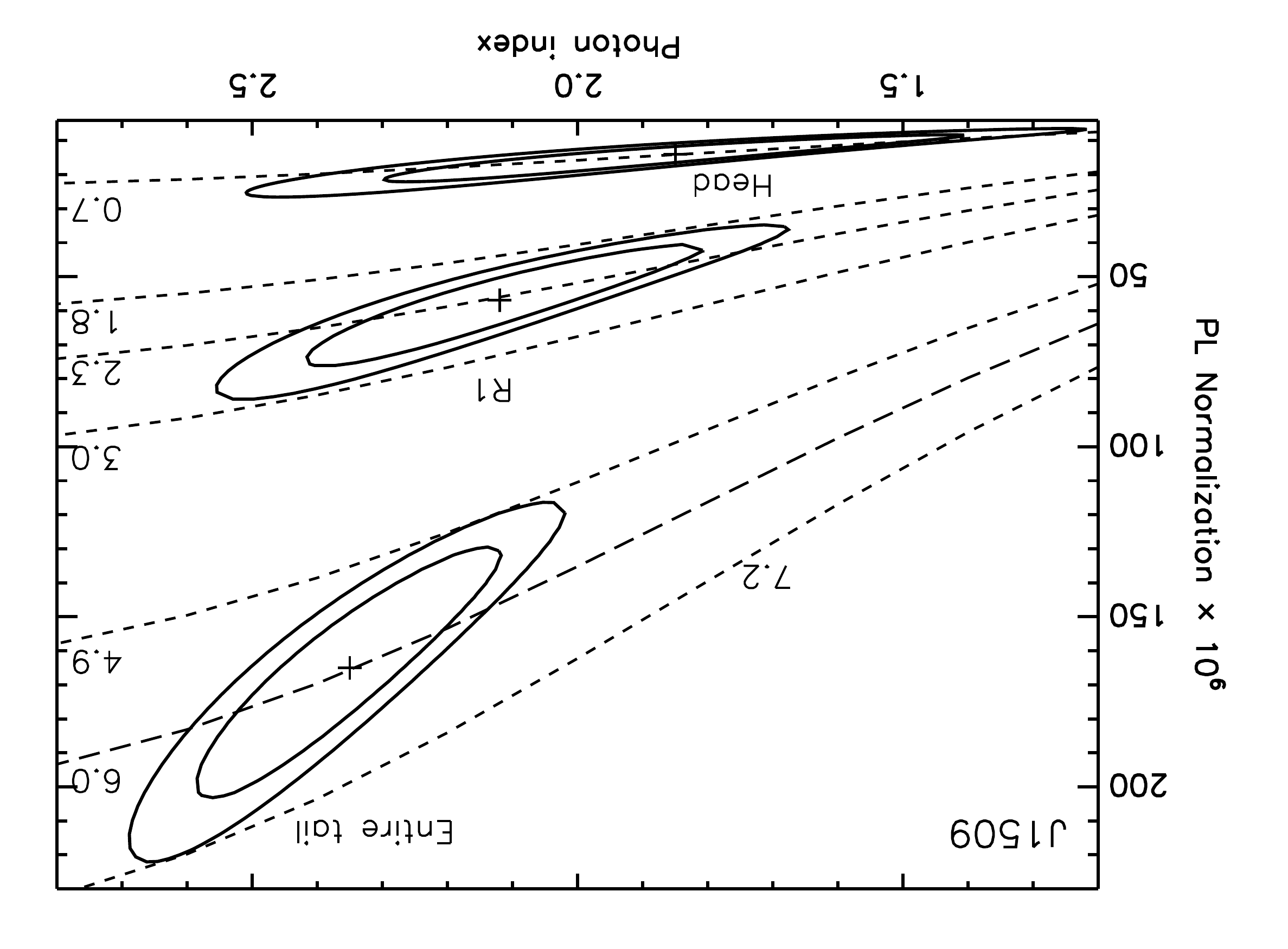}
\caption{Confidence contours (68\% and 90\%) for the PL fits to the
 J1509 PWN regions (labeled in the plot)  for the
 fixed $n_{\rm H,22}=2.1$.
 The PL normalization
is in units of $10^{-6}$ photons cm$^{-2}$ s$^{-1}$ keV$^{-1}$ at 1
keV. The dashed curves are the loci
 of constant unabsorbed flux in the 0.5--8 keV band;
the flux values near the curves are in units of 10$^{-13}$ ergs
cm$^{-2}$ s$^{-1}$.
\label{1509-contours2} }
\end{figure}

 With the fixed $n_{\rm H,22}=2.1$, the absorbed PL fits
to the R1 spectrum and the entire tail spectrum  result in the photon indices
  $\Gamma_{\rm R1}=2.1\pm0.2$ and   $\Gamma_{\rm tail}=2.4\pm0.15$, with
  the unabsorbed 0.5--8 keV fluxes
$F_{\rm R1}^{\rm unabs}=
2.33^{+0.27}_{-0.22}\times10^{-13}$ ergs cm$^{-2}$ s$^{-1}$ and
$F_{\rm tail}^{\rm unabs}=(5.9\pm0.5)\times10^{-13}$ ergs cm$^{-2}$ s$^{-1}$
 (see also Fig.\ \ref{1509-contours2} and
   Table \ref{spectral-1509}).
 The unabsorbed
0.5--8 keV luminosity of the entire tail is $L_{\rm tail}\approx1.1 \times
10^{33}d_{4}^{2}$ ergs s$^{-1}$, while the luminosity of the head  is
 a factor of 10 lower
(see Table~\ref{spectral-1509} for other details).

  The pulsar's spectrum will be analyzed in detail elsewhere.
  For the purposes of this paper, it suffices to say
  that
  the spectrum  fits the absorbed PL model with $\Gamma_{\rm psr}\simeq2.2$
 and an unabsorbed  0.5--8 keV flux $F_{\rm psr}^{\rm unabs}
\approx7.8\times 10^{-14}$
  ergs s$^{-1}$ cm$^{-2}$ (for  $n_{\rm H,22}=2.1$),
 corresponding to the isotropic luminosity
$L_{\rm psr}=1.5\times10^{32} d_4^2$ ergs s$^{-1}$.

\subsubsection{J1740}

 Since the ACIS imaging exposure of J1740  was very short, we
 have to rely upon the {\sl XMM-Newton} data to
infer the spectral properties of the
 extended tail. We excluded  events with energies below 0.4 keV and
above 10 keV from the analysis
because of the large particle background contribution at these energies.
   We then extracted the  spectrum of the bright part of the tail
(within the PN FOV)
from the
 elliptical region shown in Figure~\ref{tail-1740-mos-pn}.
 The choice of the background region
 was limited by the size of the PN
FOV in the Small Window mode,
so we extracted
it from the largest possible circular region, also shown
 in Figure~\ref{tail-1740-mos-pn}.
 In the first observation
we collected 1459, 471, and 492  counts (of which  25.6\%, 38.0\%, and 27.9\% counts come from the source) in PN, MOS1, and MOS2, respectively.
A slightly smaller numbers of counts were collected from the second
observation: 986 (28.4\% from the source) in PN, 342 (46.1\%) in
MOS1, and 353 (44.9\%) in MOS2.

\begin{figure}
\begin{center}
\includegraphics[height=12.0cm,angle=0]{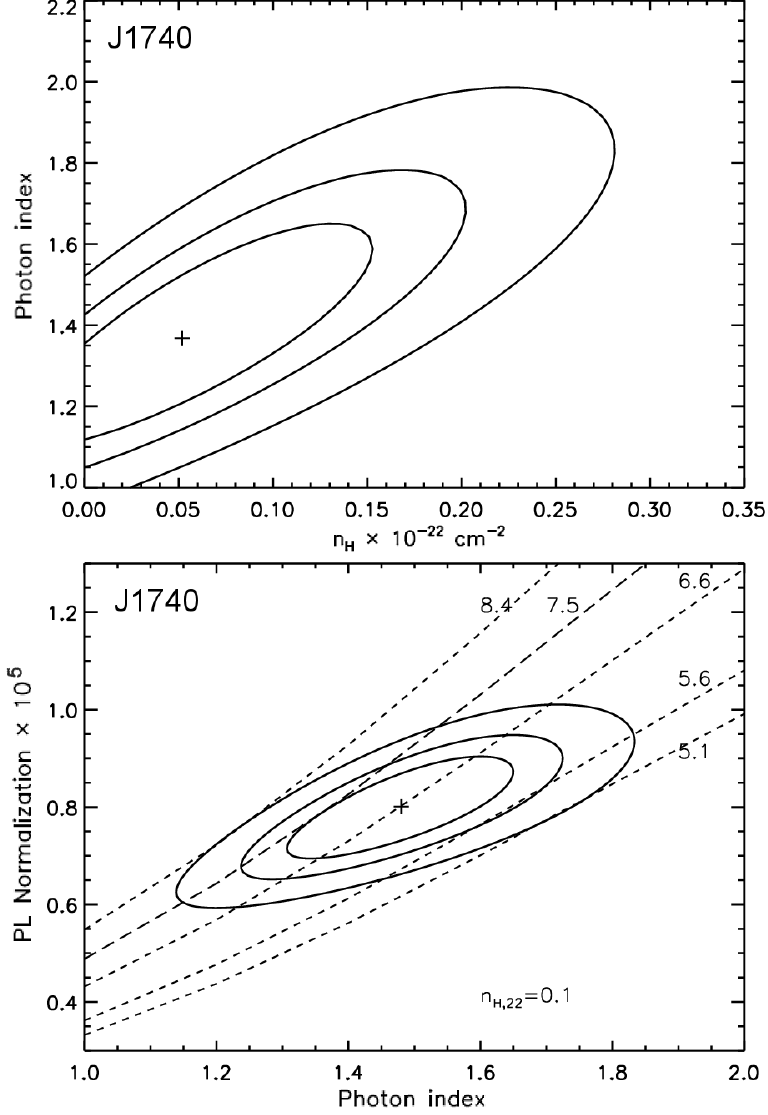}
\end{center}
\caption{
{\em Top:}
 Confidence contours (68\%, 90\%, and 99\%) in the $n_{\rm
H}$--$\Gamma$ plane for the PL fit to the J1740 tail spectrum.
The contours
 are obtained with the PL normalization fitted at each point
of the grid.
 {\em Bottom:}  68\%, 90\%, and 99\% confidence contours for the  fixed
$n_{\rm H}=1\times10^{21}$\,cm$^{-2}$. The dashed lines are the loci
of constant unabsorbed flux in the 0.4--10\,keV band;
the flux values are in units of
 $10^{-14}$\,ergs\,cm$^{-2}$\,s$^{-1}$.
\label{1740-contours}}
\end{figure}

We fit the PN and MOS spectra of both observations
simultaneously
 in the 0.4--10\,keV band using an absorbed PL model.
 First, we fit the data  with all parameters allowed to vary.
The hydrogen column density obtained from this fit is poorly
constrained, $n_{\rm H,22}=0$--0.16 (see Fig.\ \ref{1740-contours}, {\em top},
and Table 3).
Then, we fixed the absorption column to the value
 $n_{\rm H,22}= 0.1$,  consistent with the pulsar's
  dispersion measure
assuming the 10\% ISM ionization
and the fits of the pulsar spectrum (Misanovic et al., in prep.).
The results of the spectral fitting are shown in
Table \ref{spectral-1509}
 and Figure \ref{1740-contours} ({\em bottom}).
The energy flux from the tail is $F_{\rm tail}=(6.0\pm0.6)\times10^{-14}$  ergs cm$^{-2}$ s$^{-1}$ in the 0.4--10 keV band.
The tail spectrum
is relatively hard, with the photon index
 $\Gamma_{\rm tail}=1.1$--1.7; its unabsorbed  flux,
 $F_{\rm tail}^{\rm unabs}\approx
6.6\times 10^{-14}$ ergs s$^{-1}$ cm$^{-2}$,
corresponds to the 0.4--10\,keV luminosity $L_{\rm tail}
\approx 1.5\times 10^{31} d_{1.4}^2$ ergs s$^{-1}$
($L_{\rm tail}=
1.3\times10^{31} d_{1.4}^2$\,ergs s$^{-1}$ in the 0.5--8 keV band).

  The MOS and ACIS images (Figs.\ \ref{tail-1740} and \ref{1740-ACIS})
suggest that the tail extends beyond the elliptical region
   used in the above spectral analysis,
up to $\sim 7'$ from the pulsar.
However, poor statistics precludes meaningful  analysis of the
    spectra extracted outside the PN FOV
 A crude estimate of the remaining tail flux can be obtained from
    net tail counts measured in the MOS and ACIS images
outside the elliptical region.
Using  PIMMS\footnote{http://heasarc.gsfc.nasa.gov/Tools/w3pimms.html}
and assuming $n_{\rm H,22}=0.1$ and the PL slopes inferred above,
we estimated the luminosity of the
  entire
visible tail to be
$L_{\rm tail} \sim (2$--$5)\times 10^{31} d_{1.4}^2$ ergs s$^{-1}$.

 The detailed phase-resolved analysis of the
pulsar spectrum
shows that it includes both thermal and non-thermal
emission components
(the results will be reported in a subsequent paper;
Misanovic et al., in prep.).
  Here we just note that the non-thermal part of the spectrum fits
a PL model with $\Gamma_{\rm psr}\simeq1.2$--1.4
and
luminosity
$L_{\rm psr}\sim3\times10^{30} d_{1.4}^2$ ergs s$^{-1}$ in
the 0.5--8 keV band.

\section{Discussion}

 The very long tails of the
J1509 and J1740 pulsars
are  extreme examples of ram-pressure
 confined outflows from supersonically moving pulsars. To date,
about 13  extended X-ray
sources likely associated with
fast-moving pulsars
have been found,
  although
 in some cases the interpretation of the observed elongated structures
 as  tails is tentative
(see KP08 for a review).
   The sample of well-established X-ray PWNe
 with long ram-pressure confined tails is still very limited
(see \S\,3.3 and Table \ref{pulsar-tails}).
   We expanded the
    sample by adding the two pulsars, J1740 and J1509,
  with remarkably long, prominent X-ray tails.

Despite the
 similarities in their appearance,
  pulsar tails
 show significant differences in their X-ray spectra,
luminosities,
apparent lengths, surface brightness distributions, and
multiwavelength properties.  Below we
 derive  physical properties of
  the J1509 and J1740 tails, compare them  with
  model predictions, and discuss the
physical processes
that determine the tails'
 appearance and spectra.
We also provide a comparison with other pulsar tails,
 stressing the differences
 and discussing their possible origin.

 \subsection{Morphology
of the tails and pulsar speeds.}

The appearance of the X-ray PWNe around  J1509 and J1740
  clearly shows that the pulsars move
 with   supersonic
 speeds. If the pulsar speed is supersonic,
the shocked
pulsar wind (PW) ahead of the pulsar
  is confined between the
bow-shaped termination shock (TS) and the contact discontinuity (CD) surface,
while the shocked ambient matter is confined between the CD surface
and the forward shock (FS).
 The PWN behind the pulsar
acquires
an elongated shape stretched in the direction
opposite to the direction  of the pulsar's
   motion.
 Recent numerical simulations by B05 show that
for the idealized case of an isotropic PW\footnote{
If the PW is mostly confined in the equatorial plane (i.e.,
 in the plane perpendicular to the pulsar's spin axis),
as we see in young pulsars, then the TS shape
   may strongly depend on the angle between the pulsar's velocity
  and the spin axis.}
the TS acquires a bullet-like shape, with the
  TS apex
(bullet head)
at the distance
\be
R_h \approx \left[\frac{\dot{E} }{4\pi c\, (p_{\rm
amb}+p_{\rm ram})}\right]^{1/2}
 \ee
from the pulsar,
where the PW pressure,
$p_{w} \approx \dot{E}  (4\pi c r^2)^{-1}$,
is balanced by the sum of the ambient pressure,
$p_{\rm amb}$,
 and the ram pressure, $p_{\rm ram}= \rho v^2 =
1.67\times 10^{-10} n v_7^2$ dyn cm$^{-2}$
(here
 $v_7 = v/10^7\,{\rm cm\, s}^{-1}$ is the pulsar speed,
 and $n \equiv \rho/m_{\rm H}$ is
 in units of cm$^{-3}$). Assuming $p_{\rm
 ram}\gg p_{\rm amb}$ [equivalent to ${\mathcal M} \gg 1$, where
 ${\mathcal M}=v/c_s=(3p_{\rm ram}/5p_{\rm amb})^{1/2}$
is the Mach number]
we obtain
\be
R_{h}\approx4.0\times 10^{16}\, \edot_{35}^{1/2} n^{-1/2} v_7^{-1}\,\, {\rm cm},
\ee
or, equivalently, $p_{\rm ram}\approx2.65\times 10^{-9} \edot_{35} R_{h,16}^{-2}$
dyn cm$^{-2}$,
where $\edot_{35}=\edot/(10^{35}\,{\rm ergs\,s}^{-1})$,
$R_{h,16}=R_h/10^{16}\,{\rm cm}$.

For large Mach numbers
and small values of the magnetization parameter $\sigma$ of the
pre-shock pulsar wind,
 the TS bullet's
cylindrical radius is $r_{\rm TS}\sim R_{h}$ , and the distance of its
back surface from the pulsar is $R_{b}\sim 6R_{h}$ (B05).
The numerical simulations (B05) and analytical models (Romanova et al.\ 2005)
 also show that an  extended
tail
 forms behind
 the back surface of the bullet\footnote{Note that
the numerical solutions
by B05 do not extend
beyond $\sim10R_{h}$ from the
pulsar.}.
According to B05, the flow speed within this
 tail ranges from (0.8$-$0.9)$c$
 in its sheath
 to (0.1$-$0.3)$c$ in the narrow inner channel of the tail.

For J1509, the
 $~3''$-long
 region
 of enhanced
 surface brightness immediately behind the pulsar
(see Fig.~\ref{tail-vicinity-1509}, {\em top})
 can be interpreted as emission from the TS bullet,
implying the distance  $R_{h}\sim R_{b}/6
\sim 3\times 10^{16}\,(d_4/\sin i)$ cm
to the (unresolved) TS apex ($i$ is the angle between the pulsar
velocity and the line of sight).
This assumption and equation (2)
give
$p_{\rm ram}\approx1.5\times 10^{-9}(d_4/\sin i)^{-2}$ dyn cm$^{-2}$
and an estimate for the pulsar speed as a function of ambient
number density,
\be
v\approx 400\,\edot_{35}^{1/2}n^{-1/2}R_{h,16}^{-1}\,\,{\rm km\,\,s}^{-1}
\sim 300\, n^{-1/2} (d_4/\sin i)^{-1}\,\, {\rm km\,\,s}^{-1}.
\ee
Alternatively,
 we can estimate the Mach number
  as a function of ambient pressure,
\be
\mathcal{M}
 \approx (3 \dot{E}/20\pi
R_{h}^{2}c p_{\rm amb})^{1/2}\sim
30\, (d_4/\sin i)^{-1}
p_{\rm amb,-12}^{-1/2},
\ee
where $p_{\rm amb,-12}=
p_{\rm amb}/10^{-12}\, {\rm dyn\, cm}^{-2}$
($p_{\rm amb,-12}\sim 1$ is a typical ISM pressure
in the Galactic plane; e.g., Ferri\`{e}re 2001).

 Unfortunately, the  poor angular
 resolution of {\sl XMM-Newton} and the
shortness of the {\sl Chandra} exposure
 do not allow us to constrain the size of the TS bullet  around J1740.
 However, the {\sl Chandra} image of the vicinity of J1740
(see the inset in Fig.\ \ref{1740-ACIS}) shows a brightening
  $\simeq0.7''$ ahead of the moving pulsar.
One can speculate that the brightening  corresponds to
the TS apex,
which gives $R_{h}
\sim 1.5\times 10^{16} (d_{1.4}/\sin i)$ cm.
This corresponds to the pulsar speed
\be
v \sim 400\, n^{-1/2} (d_{1.4}/\sin i)^{-1}\,\,{\rm km\,\,s}^{-1}
\ee
and the Mach number
\be
\mathcal{M}\sim
130 (d_{1.4}/\sin i)^{-1} p_{\rm amb,-13}^{-1/2}
\ee
(notice that we changed the scaling here to account for the lower
pressure at large distances from the Galactic plane).
 The estimated pulsar speed is
consistent with
$v_{\perp}\sim 200$ km s$^{-1}$,
 inferred from the interstellar scintillation measurements
by McLaughlin et al.\ (2002), if $n\sim 4d_{1.4}^{-2}(\sin i)^4$ cm$^{-3}$.
It can be reconciled with the space-averaged density
 $n\sim 0.04$--0.05 cm$^{-3}$, expected at
$z\sim 400$--500 pc (Fig.\ 2 in Ferri\`{e}re 2001), if
$\sin i \sim 0.3 d_{1.4}^{1/2}$,
which would correspond to $v\sim 600 d_{1.4}^{-1/2}$ km s$^{-1}$,
$\mathcal{M}\sim 40 p_{\rm amb,-13}^{-1/2} d_{1.4}^{-1/2}$,
and the tail's length $l\gtrsim 6 d_{1.4}^{1/2}$ pc.

As the ISM pressure is much more uniform than the density,
the Mach number values (eqs.\ [4] and [6]) are more certain
than the speed values in equations (3) and (5).
On the other hand, the ambient pressure in the pulsar
vicinity may be higher than the typical ISM pressure because the
pulsar's UV emission heats and ionizes the surrounding medium
(e.g., Bucciantini \& Bandiera 2001) . Therefore,
the Mach numbers can be somewhat lower than the numerical values
in equations (4) and (6).
Also, we should not forget
that our estimates
are based on the
assumption of isotropic distribution of the unshocked pulsar wind.
The estimates may change considerably if the wind is essentially
anisotropic, but no
 models have been published for
this case so far.

\subsection{
Magnetic field and flow speed}

Assuming some value for the ratio
$k_m=w_{\rm mag}/w_{\rm rel}$ of the magnetic energy density,
$w_{\rm mag}= B^2/(8\pi)$, to the energy density of relativistic
particles, $w_{\rm rel}$,
 the magnetic field in the tails can  be
estimated from the measured
synchrotron surface brightness
(Pavlov et al.\ 2003):
\begin{equation}
B=7.2\,\left\{\frac{k_m
}{a_p (3-2\Gamma)}
\left[E_{{\rm M},p}^{(3-2\Gamma)/2} - E_{{\rm m},p}^{(3-2\Gamma)/2}\right]
\frac{\mathcal{B}_{-8}}{\bar{s}_{17}}\right\}^{2/7}\,\,\mu{\rm G}\,.
\end{equation}
Here $\mathcal{B}=\mathcal{N}/A
= 10^{-8} \mathcal{B}_{-8}$
photons (s\,cm$^2$\,keV\,arcsec$^2$)$^{-1}$ is the average spectral
surface brightness at $E=1$ keV (see Table \ref{spectral-1509}),
$\mathcal{N}$
is the normalization of photon spectral flux
measured in area $A$, $\bar{s}=10^{17}\, \bar{s}_{17}$ cm is the
average length of the radiating region along the line of sight,
$E_{{\rm
m},p}=E_{\rm m}/y_{{\rm m},p}$, $E_{{\rm M},p}=E_{\rm M}/y_{{\rm
M},p}$, $E_{\rm m}$ and $E_{\rm M}$ are the lower and upper
energies of the photon power-law spectrum (in keV), and $y_{{\rm
m},p}$, $y_{{\rm M},p}$ and $a_p$ are the numerical coefficients
whose values depend on the slope
$p=2\Gamma-1$ of the electron power-law spectral
energy distribution (see Table \ref{synch-table}
 and Ginzburg \& Syrovatskii 1965).

\begin{table}[]
\caption[]{Synchrotron parameters and equipartition field} \vspace{-0.5cm}
\begin{center}
\begin{tabular}{lccccccrcc}
\tableline\tableline
Region & $\Gamma$ & $p$   &
 $y_{{\rm m},p}$  &
 $y_{{\rm M},p}$   &  $a_p$ & $s_{\rm 17}$ & $\mathcal{B}_{-8}$ & $B_{\rm eq,-6}$ \\
\tableline
      \tableline
&&&&J1509&&&&\\
\hline
Head & 1.8             &  2.6            &   2.3       &   0.11        &  0.082     & 5    &  5.3    &   22 \\
R1     & 2.1             &  3.2            &    2.9      &    0.22       & 0.072      & 7     &   1.4   &   16 \\
R2     & 2.4             &  3.8            &    3.3      &    0.33       & 0.071      & 20   &   1.0     &   14  \\
R3     & 2.4             &  3.8            &    3.3      &    0.33       & 0.071      & 20   &   0.6   &   12  \\
      \tableline
&&&&J1740&&&&\\
\hline
Bright portion   & 1.4    &  1.8   &    1.6      &    0.023   & 0.114     & 10   &   0.1      &      5.4  \\

       \tableline
\end{tabular}
\end{center}
\tablecomments{Parameters used to estimate the
equipartition magnetic field, $B_{\rm eq} = B_{\rm eq,-6}\,\mu$G,
from the measured surface brightness, assuming
$E_{\rm m}=0.1$ keV and  $E_{\rm M}=10$ keV (see Table \ref{spectral-1509}
and \S3.2).
 The numerical coefficients $y_{{\rm
m},p}$, $y_{{\rm M},p}$ , and $a_p$ are estimated using Table II
in Ginzburg \& Syrovatskii (1965).
 }
\label{synch-table}
\end{table}

The quantities $\mathcal{B}$ and $\Gamma$ have been directly measured
 (Table \ref{spectral-1509}), and $\bar{s}$ can
be estimated assuming that it is close to the observed tail's width.
In addition to these parameters,
the strength of the magnetic field depends on
the boundary
energies of the photon power-law spectrum, which are
rather uncertain because of the lack of
 sensitive high-resolution observations outside the X-ray range. However, the
observed softening of the J1509 tail spectrum
with increasing distance from the
pulsar (Table \ref{spectral-1509}), as well as the decrease of the tail's
length at higher photon energies
(Fig.\ \ref{1509-energy-resolved}),
 suggest that, at least for the J1509 tail, $E_{M} $ is
close to
the upper energy of
the ACIS band, $\sim 10$ keV.
 The lower energy of the synchrotron spectrum
  is not constrained; we can only be sure that $E_m$ is smaller than
the lower energy of the ACIS band, $\sim 0.3$ keV.
In the numerical estimates (Table \ref{synch-table}),
we will assume $E_{m}=0.1$ keV and $E_{M}=10$ keV.

Under these assumptions,
we obtain the equipartition magnetic fields
$B_{\rm eq}\equiv B k_m^{-2/7} \sim 10$--20 $\mu$G
in several regions of the J1509 tail,
with a hint of a decrease of $B_{\rm eq}$ with increasing
distance from the pulsar (see Table \ref{synch-table}).
For the bright portion of J1740 tail,
where the spectrum was measured, we found $B_{\rm eq}\sim 5$ $\mu$G.

We stress that the magnetic fields in Table \ref{synch-table}
may differ by a factor of a few from the actual magnetic fields.
First of all, if the energy range of the power-law spectrum
is much broader than the 0.1--10 keV band, the magnetic field
is larger than the estimates in Table \ref{synch-table}.
For instance, as $\Gamma > 1.5$ ($p>2$) for the spectra of
 the J1509 tail, the derived magnetic field is
sensitive to
the unknown $E_m$ value: $B\propto E_m^{-(2\Gamma -3)/7}$.
If the radio emission
in the SUMSS\footnote{Sydney University Molonglo Sky Survey} image
   reported by HB07 (see also Fig.\ \ref{radio-xray-color}) indeed
belongs to the J1509 tail,
then the X-ray power-law spectrum might be extrapolated
 down to $E_m\sim 10^{-4}$--$10^{-2}$ keV (depending on the $\Gamma$
value), which would correspond to a factor of 1.5--6 higher magnetic
fields. On the other hand,
the magnetic field strength may be lower than $B_{\rm eq}$ if
$k_m < 1$ (e.g., by a factor of 1.9 if $k_m= 0.1$.)
Overall,
it seems reasonable to conclude that the magnetic fields in the
J1509 tail are in the range $\sim 10$--100 $\mu$G
(probably closer to the upper end of this range), while the magnetic
field in the J1740 tail is a factor of a few lower.
To obtain more certain estimates, deeper ACIS observations and
 observations outside the X-ray range would be most important.

Using the magnetic field estimates, we can constrain the maximum
electron Lorentz factor required for producing the observed
X-ray power-law spectrum: \\
 $\gamma_M \approx 7.6\times 10^7
(E_{\rm M}/y_{{\rm M},p})^{1/2} (B/10\,\mu{\rm G})^{-1/2}
\gtrsim 3\times 10^8$ (i.e., the
maximum electron energy $E_{e,{\rm M}}\gtrsim 150$ TeV),
 for the J1509 tail\footnote{We should note that HB07 assumed
$\gamma=10^6$ and derived the magnetic field $B\sim 5$--7 $\mu$G
in the shocked region. Such parameters are inconsistent with the X-ray data
because they correspond to the characteristic
energy of synchrotron radiation, $E\sim 0.03$ eV, in the infrared range
($\lambda \sim 30\,\mu$m), five orders of magnitude below the
observed X-ray energies.}.
 The Larmor radius for such electrons,
$r_L = 1.7\times 10^{8}
\gamma (B/10\,\mu{\rm G})^{-1}\,\,{\rm cm}$,
is $ \gtrsim 4\times 10^{16}$ cm;
it is smaller than the observed tail width as long as $\gamma_M \lesssim
10^{10}$.

The magnetic fields derived above
and the measured projected tail length  allow one to constrain the
 average projected flow
speed in the tail:
\be
v_{\rm flow,\perp}\sim l_\perp/\tau_{\rm syn}
\sim 860\, (l_\perp/1\,{\rm pc}) (B/10\,\mu{\rm G})^{3/2}
(E/1\, {\rm keV})^{1/2}\,\,{\rm km}\,\,{\rm s}^{-1},
\label{flow-speed}
\ee
where
\be
\tau_{\rm syn}=
5.1\times 10^8 \gamma^{-1} B^{-2}\,\,{\rm s}
\sim 1100 (B/10\,\mu{\rm G})^{-3/2} (E/1\, {\rm keV})^{-1/2}\,\,{\rm yr}
\label{tau-syn}
\ee
is the synchrotron cooling time [here we used the characteristic
Lorentz factor, $\gamma \sim 1.4\times 10^8 (B/10\,\mu{\rm G})^{-1/2}
(E/1\, {\rm keV})^{1/2}$ for the electrons that give the main contribution
to the synchrotron radiation at energy $E$].
Using equation (\ref{flow-speed}), the magnetic fields in Table
\ref{synch-table}, and the highest energies at which the different
regions of the J1509 tail are seen, we obtain $v_{\rm flow,\perp}\gtrsim
7,500$ km s$^{-1}$ for region R1, and $v_{\rm flow,\perp}\sim 13,000$
and 12,000 km s$^{-1}$ for regions R1+R2 and R1+R2+R3, respectively,
for $d=4$ kpc.
Given the uncertainty of the magnetic field estimates, the actual
average flow speeds can be in the range $v_{\rm flow,\perp}
\sim 5,000$--100,000
km s$^{-1}$, with higher values more plausible than lower ones.
A similar consideration for the J1740 tail yields $v_{\rm flow,\perp}
> 1000$ km s$^{-1}$, but we should note that this is only a conservative
lower limit (because the actual projected tail length is likely
larger than observed), and $v_{\rm flow}=v_{\rm flow,\perp}/\sin i$
can substantially exceed the projected velocity for small $\sin i$.
More certain estimates for $v_{\rm flow}$ could be obtained by
direct measurements of the proper motion of the emission ``clumps''
(local brightenings) in the J1509 tail (see Fig.\ 5, {\em bottom left}),
similar to the proper motion
 measurements of moving ``blobs'' in the Vela pulsar
jet (Pavlov et al.\ 2003), which would require a series of deep
ACIS observations.

Even with account for the large uncertainty, we can conclude that,
at least for the J1509 tail, the
flow speed is much larger than any reasonable pulsar speed.
This means that the equation $\tau_{\rm syn}=l/v_{\rm psr}$,
used in many papers on  bowshock-tail
PWNe\footnote{Including HB07.},
is inapplicable. (This equation implies that the downstream flow
instantly acquires the velocity of the ambient medium [i.e.,
$v_{\rm flow}= v_{\rm psr}$ in the pulsar reference frame],
leaving a synchrotron-emitting ``trail'' behind the moving pulsar.)
On the other hand, the inferred flow speed is
smaller than $v_{\rm flow}
\sim 0.8c$--$0.9c$ obtained in numerical simulations by B05 for
the bulk of the tail volume behind the TS bullet. However, those simulation
were done only for the beginning of the tail (e.g., for distances
$\lesssim 13 R_h$ from the pulsar in B05), whereas we analyze a much
larger extent of the tail (up to $l_\perp \sim 600 R_h$).
 It seems
plausible that the flow is
mildly relativistic in the immediate vicinity
behind the TS bullet, and it decelerates at larger distances because of the
 expansion of the overpressurized flow,
entrainment of the ISM material (e.g., caused by the shear instability at the CD surface separating the fast-moving
shocked PW from the shocked ambient ISM), or internal shocks in
the tail.

The broadening
 of the J1509 X-ray tail
with increasing distance $z$ from the pulsar (Fig.\ \ref{1509-profile2})
 suggests that
the cross-sectional area of the tail
increases up to $z\approx 1.5'$, as $S(z)\propto z^\kappa$,
with $\kappa\simeq0.7-1.5$.
 At larger distances, $2'\lesssim z <\lesssim 5'$,
   the tail seems
  to have more or less constant width,
although this requires confirmation with deeper
  observations. The observed
radial expansion of the tail at large $z$
 significantly exceeds the predictions by
  Romanova et al.\ (2005; eqs.\ [32--34]), for plausible values of
$v_{\rm flow}$ and $v_{\rm psr}$ (see Fig.\ \ref{wilkin}).
 Including the interaction between the shocked pulsar wind
and shocked ambient gas at the CD surface might help to bring the model
predictions in agreement with the observations.
Polarimetric observations in the radio would allow one
to measure the direction of the magnetic field in the tail and test whether
 it is consistent with the toroidal field assumption adopted in the current
models\footnote{So far, only the Mouse PWN tail
has been studied in the radio,
 and
the direction of the detected polarization
suggests that the magnetic filed
is predominantly aligned with
tail direction (Yusef-Zadeh \& Gaensler 2005), contrary
to the current models.}.
We should note that the relationship between $B$, $S$, and $v_{\rm flow}$
at given $z$ depends on the magnetic field topology. For instance,
in ideal MHD models,
$B\propto v_{\rm flow}^{-1} S^{-1/2}$ for purely toroidal magnetic
field (B05), while $B\propto S^{-1}$ if the magnetic field is
parallel to the tail direction.
Deep X-ray observations would allow one to
measure both $S(z)$ and $B(z)$ and test the validity
of the ideal MHD approximation and the impact of the ambient matter
entrainment.

\subsection{Multiwavelength Aspects}

\begin{figure}[t]
 \centering
\includegraphics[width=3.2in,angle=0]{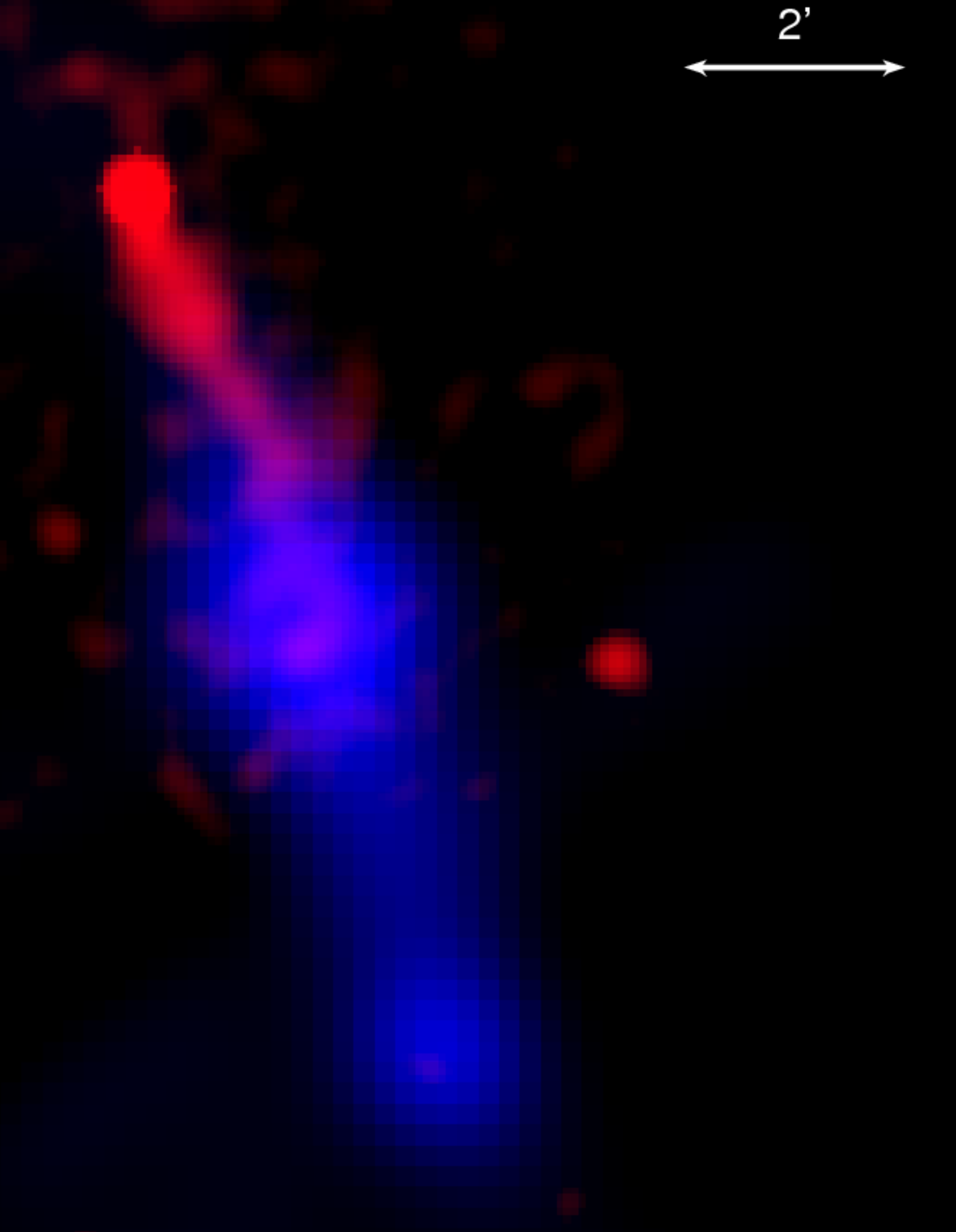}
 \caption{ Composite X-ray (red) and radio (blue)
 image of the field of the J1509 tail.}
 \label{radio-xray-color}
\end{figure}

The good alignment between the X-ray tail of J1509 and
 the elongated radio feature seen in the SUMSS image (Fig.\
\ref{radio-xray-color})  suggests
 their physical association (HB07).
As it was  noted by HB07,
the larger length of the J1509 radio tail
could be explained by slower synchrotron cooling of the radio-emitting
electrons.
However, the elongated radio feature has highly non-uniform
surface brightness,
with two  maxima.
The maximum
located at $\approx8'$  from the pulsar
 is clearly associated with an
unrelated point source, possibly an AGN.
 (The source is also seen in X-rays; it is marked ``R'' in
Fig.\ \ref{tail-1509}.)
 The
origin of the other maximum
(located about $5'$ from the pulsar) is
 unclear. It appears to be
extended and could be a remote SNR (as suggested by Whiteoak \&
Green 1996), a radio galaxy, or an inhomogeneity in the pulsar wind
flow
 (e.g., due to an internal shock or instability).
 To discriminate between these possibilities,
  deeper radio and X-ray observations are needed.

Another puzzling aspect is the gradual
 decrease in the radio surface
brightness of the tail toward the pulsar (the brightness drops by at least a
factor of 5 from its maximum value at $\approx5'$ from the pulsar),
in contrast with the
X-ray surface brightness that increases toward the pulsar (see
Fig.\ \ref{radio-xray-color}).
 Unfortunately, the SUMMS image is not deep
enough to show whether or not the putative radio tail actually connects to
the pulsar.
  If the radio emission is indeed associated with the J1509 tail, the decrease in the radio surface brightness
 toward the pulsar
  cannot be
  explained by the synchrotron cooling
because the corresponding cooling time,
$\tau^{\rm radio}_{\rm syn}=1.7\times10^{7}(B/10\mu\rm{G})^{-3/2}
 (\nu/1{\rm
GHz})^{-1/2}$ yrs, greatly exceeds the
pulsar's age.
   To explain the lack of  radio emission closer to the pulsar, one
  has to assume that
the number density of particles, $n(z)$, increases with the distance
from the pulsar [e.g., $n(z)\propto 1/v_{\rm flow}(z)$ for a constant
cross-sectional area of the tail).
  This increase would not result in higher X-ray surface brightness
if the most energetic  electrons, responsible for the X-ray emission,
  undergo strong synchrotron cooling,
  such that their emission moves out of the
X-ray band by the time
when the flow reaches the distance where $n(z)$ increases substantially.
 Similar spatial radio-X-ray  anti-correlation
 has been reported  for the PWNe around
the Vela pulsar (Dodson et al.\ 2003; Kargaltsev \& Pavlov 2004)
and PSR B1706$-$44 (Romani et al.\ 2005),
which are moving with $\mathcal{M}\lesssim 1$ .
  On the other hand, in the Mouse PWN
   (which is the only
    pulsar tail well studied in the
radio and X-rays)
both the radio and X-ray emission are brightest
near the pulsar and gradually
 fade  with increasing $z$.
Note that the Mouse X-ray tail
appears to be
 shorter
($l_\perp \sim 1$ pc),
  or at least its X-ray surface brightness drops
faster with the distance from the pulsar than in the J1509 and J1740
tails,
 which could result from  faster cooling or
 lower flow
speed.
  We also note that the remarkable $12'$-long ($l_\perp = 17$ pc at $d=5$ kpc) radio tail in the Mouse
(Yusef-Zadeh \& Gaensler 2005) is a factor of $>20$ longer
than its X-ray tail, yet
 the radio tail remains well collimated and straight along its entire visible extent.
  Only deep radio observations with better angular resolution
  will allow one to understand the nature of the radio emission coincident with the
 J1509 tail and thus determine its true extent.
Confirming the radio counterpart
of the J1509  tail
  and measuring the magnetic field distribution in the radio tail
  will significantly advance our understanding of collimated magnetized outflows from pulsars.

\begin{figure}[t]
 \centering
\includegraphics[width=3.5in,angle=0]{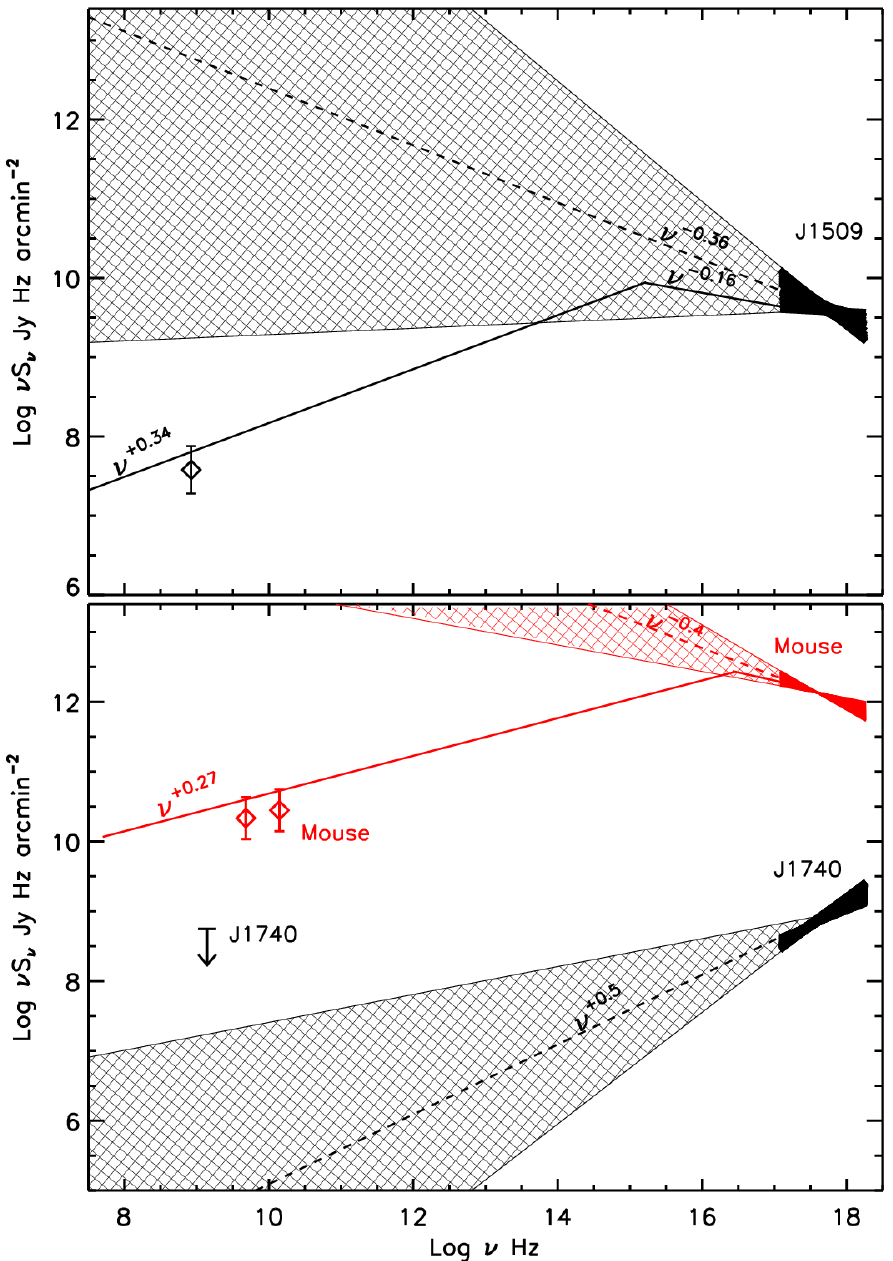}
 \caption{ Multiwavelength spectra of extended pulsar tails.
For J1509 ({\em top panel}),
 the X-ray and radio fluxes are taken from region R2
(see Fig.\ \ref{regions}).
    For J1740 ({\em bottom panel; black}),
the flux is taken from the bright X-ray tail region
(the ellipse in Fig.\ \ref{tail-1740-mos-pn}). For the Mouse
 PWN ({\em bottom panel; red})
 the X-ray and radio fluxes are for the tail region as defined in the
Figs.\ 3 and 5 of Gaensler et al.\ (2004). An upper
 limit on the J1740 1.4 GHz flux, estimated by V.\ Kondratiev (priv.\ comm.)
from archival VLA data
is also shown in the bottom panel.
The shaded areas show the uncertainties of the X-ray
 spectra extrapolation. The broken solid lines show
examples of possible  spectra
of the J1509 and Mouse tails,
which are consistent with both X-ray and radio
 measurements (the detection of the radio tail in J1509 is
tentative; see text).
The spectral breaks correspond to the change of 0.5 in the spectral index
 due to synchrotron cooling.   }
 \label{multiwave-spec}
\end{figure}

The multiwavelength  (radio through X-rays) spectra of the Mouse, J1509,
and J1740 tails are shown in Figure \ref{multiwave-spec}. If
the tentative
 interpretation of the extended radio emission
as a counterpart of the J1509  X-ray tail is correct,
 then the tail spectrum
should exhibit a spectral break between the radio and
X-ray frequencies.
Assuming that the break in the spatially integrated spectrum is due to
the synchrotron cooling ($\Delta\Gamma\sim0.5$), the break frequency
is between
 $\sim 10^{12}$
 and $\sim10^{16}$
  Hz,
for the allowed range of the X-ray spectral slopes
(an example of such a break
is shown in Fig.~\ref{multiwave-spec} for $\Gamma_X = 2.16$).
Low-frequency breaks (in a 10--100 GHz range) have
been previously reported in a number of PWNe,  such as G11.2--0.3,
G16.7+0.1, and G29.7--0.3 (Bock \& Gaensler 2005).  The spectrum of
the Mouse also shows similar behavior, indicating a break around
$10^{16}$ Hz (Fig.\ \ref{multiwave-spec}).

No extended radio emission has been reported around
J1740. However, this could be due to the fact that the field has not yet been
 observed with sufficiently deep
exposures. The X-ray spectrum of the J1740 tail is substantially harder
 than those of the J1509 and Mouse
tails, suggesting a less
efficient cooling or a different injection spectrum.
Also,
 one should keep in mind that,
 most likely,
we have detected only
 the $\sim2$ pc  brighter part of the J1740 tail,
where the cooling makes no significant effect,
while the entire tail can be
 substantially longer.
A deep {\sl Chandra} ACIS observation is necessary
  to reveal
  the true extent of the J1740 tail and search
for spectral changes associated with the cooling.

So far, no TeV emission has been found in the vicinity of
bowshock-tail X-ray PWNe,
 with the possible exception of the nebula associated with
PSR B1509--58 (Aharonian et al.\ 2005),
 where the origin of the  bright  feature
southeast of  the pulsar
(Gaensler et al.\ 2002) is not completely understood
(it could be a jet, a tail,
 or, more likely,
a combination of those,
i.e., a jet
behind the pulsar
moving at $\mathcal{M}\sim 1$).
However, the X-ray images of J1509 and J1740 reveal the presence of
relativistic electrons with $\gamma \gtrsim 10^{8}$, which,  for a
plausible magnetic field $B\gtrsim 10$ $\mu$G,
can produce
TeV emission via
the inverse Compton scattering of the
background IR and CMB photons.
 Furthermore, given the large
angular extent of these tails, their TeV emission can be resolved
with the current TeV observatories such as HESS, MAGIC and
VERITAS. The high energy GeV  emission may also be detectable with
{\sl GLAST} if there is a substantial density of synchrotron radio
photons that will be up-scattered into the GeV range.

\subsection{Comparison with other pulsar tails}

\begin{figure}[t]
 \hspace{-0.5cm}
\includegraphics[width=3.7in,angle=180]{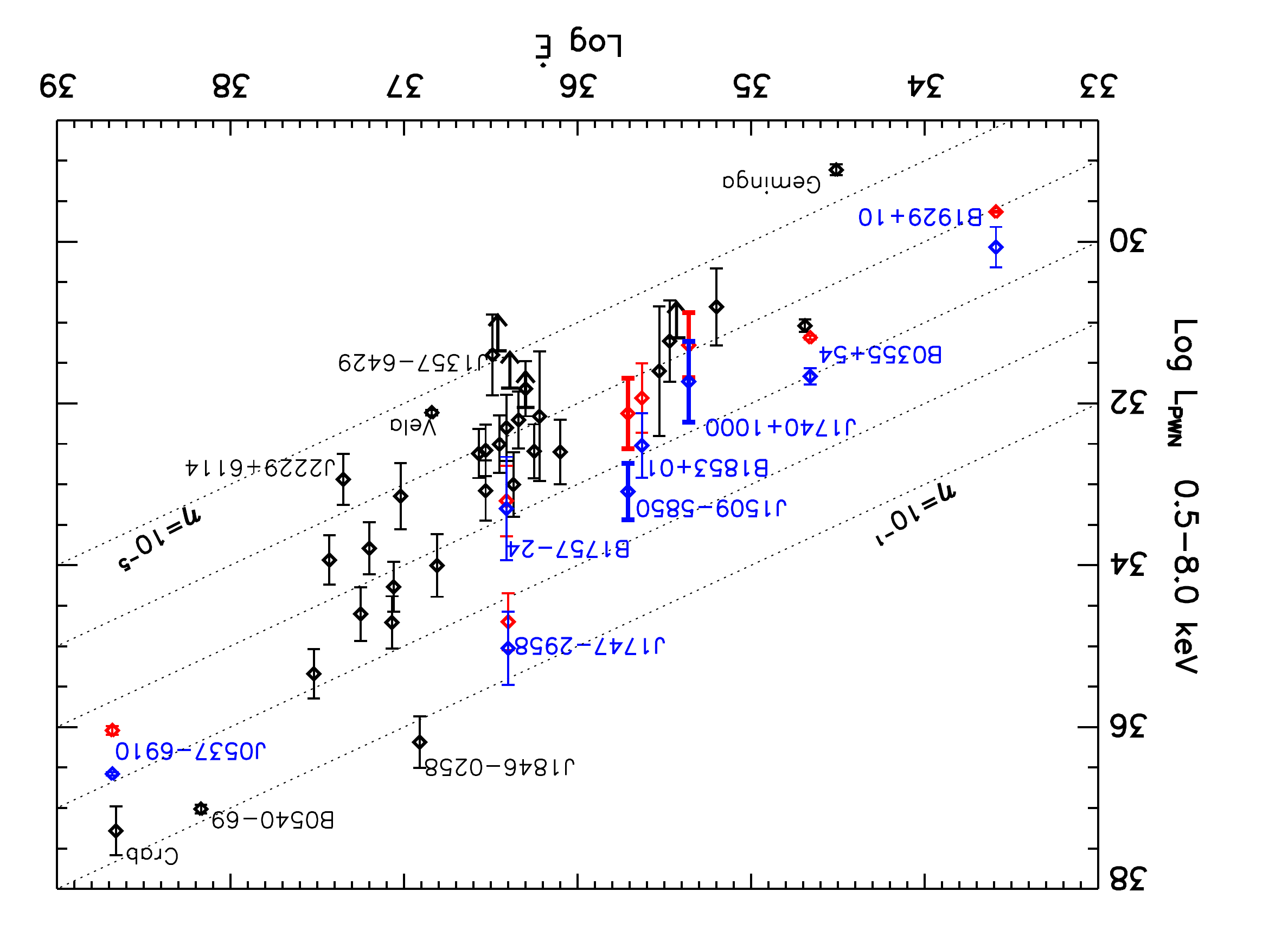}
 \caption{X-ray luminosities of PWNe versus
pulsar spin-down power (see KP08).
 The data points corresponding
  to pulsars with  firmly established long X-ray tails
  (see Table \ref{pulsar-tails}) are shown in color (as well as the names of the pulsars).
The blue and red points correspond to the total PWN luminosities
and the luminosities of the
   compact bright features  associated with the PWN ``heads'',
respectively.
In most cases
 the  luminosity of the extended tail component substantially exceeds that
of the head. }
 \label{Lx-Edot}
\end{figure}

Table \ref{pulsar-tails} summarizes the X-ray properties of
eight
PWNe with firmly established long X-ray tails.
 Two most luminous tails belong to the very young, energetic
PSR J0537--6910\footnote{
We should note, however, that, because of high pressure in the young
SNR, this pulsar may be moving subsonically, in which case the extended
structure is a ``trail'' rather than a tail
(see \S3.2), and hence it cannot be meaningfully compared with
the tails of supersonically moving pulsars.}
(located in the LMC SNR N157B)
 and the
 26 kyr old PSR J1747--5928 (the Mouse PWN), whose
X-ray efficiencies, $\eta_{\rm pwn}\equiv
L_{\rm pwn}/\dot{E}$,
  are $2.2\times10^{-2}$ and $1.6\times 10^{-2}$, respectively.
 The J1509 tail ranks third in terms of its efficiency,
 $\eta_{\rm pwn}\approx2.6\times10^{-3}$, which exceeds the
efficiency of, e.g., the
Duck PWN ($\eta_{\rm pwn}=3.2\times10^{-4}$), created by the younger and more powerful PSR B1757--24,
whose spin-down properties are very similar
   to those of PSR J1747--5928. This once again
suggests that, although there is a positive correlation
of $L_{\rm pwn}$ with $\edot$ (see Fig.\ \ref{Lx-Edot}),
 the
X-ray PWN luminosity also depends
 on other parameters (see KP08).
   The remaining four  pulsars (B1853+01, J1740+1000, B0355+54, and B1929+10)
     show similar PWN efficiencies,
 $\eta_{\rm pwn}=(2$--$7)\times10^{-4}$, despite the fact that their
spin-down ages and powers differ by more than 2 orders of magnitude.
  Although some of this scatter
could be attributed to the poorly known distances and
    different sensitivities of the individual X-ray observations,
   it is likely that, in addition to the spin-down parameters, there
    are other parameters governing
    the radiative efficiency of the winds in extended pulsar tails,
such as   the pulsar speed, the ambient pressure, and  the angles between the
NS magnetic and spin axes
 and the velocity vector.

 In addition, for the
collimated tails with nearly relativistic flow
 speeds, the orientation
 with respect to the observer
  can be important because
  the tail
brightness
is proportional to $(1+\beta\,\cos i)^{\Gamma +2}$, where
$\beta=v_{\rm flow}/c$ and $i$ is the
angle beween the velocity vector and the direction to the observer.
 For
instance, the ratio of the brightnesses at $i=30^\circ$ (approaching
flow) and $i=120^\circ$ (receding flow)
can be as large as $\simeq10$ for
$\beta=0.3$ and $\Gamma=2.3$. In addition to the
brightness changes, the  outflow components  approaching and receding
at small angles will appear shorter  simply
as a result of
 projection. This might explain why the luminous X-ray
tail of the more energetic Mouse PWN is so much shorter than that
of the less energetic J1509, or why the X-ray tail of
the Duck PWN is
not only short but also very dim compared to the X-ray tail of the Mouse.

 Comparing the X-ray luminosities  of pulsar tails with
 those of
PWNe around slowly moving
 pulsars, we see that, on average, the tails show higher X-ray
efficiencies
(see Fig.\ \ref{Lx-Edot}).
This is a natural consequence
  of the fact that in the ram-pressure confined PWNe
   the energy of the wind flow is channeled into a narrow
cylindrical
structure (the
tail). This results in a higher column density of emitting
electrons, hence
a higher surface brightness, which
   allows one to detect the wind emission
 at larger distances from the
pulsar compared to PWNe around slowly moving pulsars
   where the wind flow is more isotropic.

   If the starting flow speed
just downstream of the TS back surface
is more or less the same (mildly relativistic) for different pulsars,
 then the observed  tail
 lengths
are determined by the magnetic field strengths
in the tails (because the synchrotron cooling limits the length)
 and by the rates at which the tail widths increase
with increasing distance $z$ from the pulsar
(because the tail broadening reduces its surface brightness).
The magnetic field in the tail	
should be proportional to
$(\sigma\edot)^{1/2}$ ($\sigma$ is
 magnetization parameter in the pre-shock wind).
  The lack of correlation with $\edot$  in the current,
 very limited sample (Table 5) might be attributed to different wind magnetization in different pulsars
 and/or a large spread of the unknown inclination angles of the tails.
 The tail width is expected to show a negative correlation with the ambient pressure
(e.g., $R_{\rm tail}\propto [\edot/(v_{\rm flow}^3 p_{\rm amb})]^{1/4} z^{1/2}$ in the
Romanova et al.\ 2005 model);
 however, the dependence is too weak to be tested with the existing data.

\section{Summary}

We have presented a detailed analysis of the properties of two
PWNe associated with the supersonically moving pulsars J1509--5850
and J1740+1000. In both cases, the X-ray images
  reveal parsec-scale tails
  connected to the pulsars.
 The observed extraordinary lengths of the
tails suggest that the tail flow starts as mildly relativistic,
 and its speed remains much higher than the
 pulsar speed
 up to a distance of a few parsecs.
  The large extent of the J1509 tail suggests an average flow speed
 in excess of $\sim 5000d_{4}$ km s$^{-1}$ (a conservative lower limit).
  We estimate the average equipartition field in the J1509 tail to be   a few $\times 10^{-5}$ G,
 with a hint of weakening with the distance from the pulsar.
  The equipartition field in the J1740 tail
  is likely to be a factor of a few lower.
A noticeable difference between the two tails
 is that the spectrum of the J1740 tail is substantially harder,
implying
 that either the cooling is not as efficient in the J1740 tail
(consistent with the inferred lower magnetic field)
or the electron injection spectrum
 of J1740
 is intrinsically harder compared to that of J1509.
The projected orientation of the J1740 tail on the sky
  suggests
 that the pulsar is moving
 toward the Galactic plane. This
  means that
 the pulsar was born
well above the Galactic disk
rather than ejected from the disk with a high speed.

 The existing shallow radio survey  data suggest a possible radio
counterpart to the J1509 tail; however, a confusion with
   the background sources remains a possibility until better quality
radio data are obtained. If the radio feature is indeed
   associated with the J1509 tail, then the tail spectrum should
exhibit a break between 10$^{12}$ and 10$^{16}$ Hz.

  The comparison between the J1509 and J1740 tails and the X-ray
tails from other fast-moving pulsars shows that the X-ray luminosity
 does not correlate well with the pulsar spindown luminosity and age;
other physical parameters such as the pulsar velocity,
the ambient pressure and sound speed, and the angle between the pulsar's
 spin and magnetic axis may be equally important. In general,
we find that the X-ray efficiencies  of the ram-pressure
confined PWNe are systematically higher
  than those of PWNe around slowly moving pulsars with  similar
spindown parameters.

\acknowledgements{}

 We thank Vlad Kondratiev for the help with the archival radio data,
and Niccolo Bucciantini and Richard Lovelace for the useful discussions
of theoretical models.
 This work was partly supported by NASA grants
NNX06AH67G and NAS8-01128, and
{\sl Chandra} award SV4-74018.

\begin{table*}[]
\caption[]{Properties of long pulsar tails.} \vspace{-0.5cm}
\begin{center}
 \begin{tabular}{lcccccccccc}
\tableline\tableline PSR & SNR &$d$\tablenotemark{a} &  $v_{\perp}$\tablenotemark{b} & $\log\tau$ &  $\log\dot{E}$ & $l_\perp$\tablenotemark{c} & $\log L_{\rm pwn}\tablenotemark{d}$   &
 $\log \eta_{\rm pwn}$ & Rad.\tablenotemark{e} &  Ref.\tablenotemark{f} \\
\tableline
 &  &     kpc  & km s$^{-1}$ & [yrs]  & [ergs s$^{-1}$] & pc  & [ergs s$^{-1}$] &  & & \\
\tableline
J0537--6910        &  N157B  & 50                    &  600                     &      3.70   &   38.68 &    3.7    &     $36.43\pm0.02$ & $-2.24$ &  Y  &  1,2 \\
B1757--24         & ...                 &  5                             &   130           &   4.19  & 36.41   &   0.5      &    $33.20\pm0.14$ & $-3.21$ &  Y  & 3  \\
 J1747--2958       &  ...                &  5                             &  300           &      4.41   &   36.40 &    1.1    &     $34.75\pm0.15$ & $-1.65$ &  Y  & 4  \\
 J1509--5850       &  ...                &  4                             &  400            &      5.19   &   35.71 &    6.5    &     $33.04\pm0.05$ & $-2.67$ &  P  &  tw  \\
  B1853+01        &   W44      &  3                              &   100           &      4.31   &   35.63 &    1.3    &     $32.39\pm0.10$ & $-3.24$ &  Y  &  5  \\
J1740+1000                   &  ...                &  1.4                         &  400              &      5.06   &   35.36 &    2      &     $31.54\pm0.20$ & $-3.82$ &  N   & tw  \\
\\
 B0355+54                      &                     &  $1.04^{\rm p}$     &   $61^{+12}_{-9}$                &      5.75   &   34.66 &    1.5    &     $31.49\pm0.10$ & $-3.16$ &  N  & 6  \\
 B1929+10                      &  ...                &  $0.36^{\rm p}$      &  $177^{+4}_{-5}$                &      6.49   &   33.59 &    1.5    &     $29.87\pm0.25$ & $-3.72$ &  P & 7,8  \\
  \tableline
\end{tabular}
\end{center}
\tablecomments{The table includes pulsars with  established
parsec-scale X-ray tails.
}
\tablenotetext{a}{
Adopted distance to the pulsar (superscript $^{\rm p}$ marks the distances
obtained from parallax measurements).}
\tablenotetext{b}{Pulsar's transverse velocity, either obtained from
proper motion and parallax measurements (for B0355+54 and B1929+10;
Chatterjee et al.\ 2004)
or estimated from eq.\ (3), with $n=1$ cm$^{-3}$  and $R_h$
  measured from the
X-ray data.
The exception is J0537--6910, for which we quote the velocity
estimated by Wang et al.\ (2001) from the pulsar travel time
(this pulsar may be moving subsonically
because of the high sound speed
in the young host SNR).
Also
note that since PSR B1853+01 is  embedded in a plasma that
shows strong thermal X-ray emission, the assumption
$n=1$ cm$^{-3}$
likely overestimates $v_{\perp}$ in this case.
On the contrary, since PSR J1740+1000 is likely moving
in a rarefied medium above the Galactic plane,
its speed is
probably underestimated
(see text for discussion).
}
\tablenotetext{c}{Projected
length of the tail.}
\tablenotetext{d}{X-ray luminosity
 of the tail (including the brighter compact component
in the pulsar vicinity) in the 0.5--8 keV band
 as we measured it from {\sl Chandra} and {\sl
XMM-Newton} data.
The values may differ from those published in the
original works (see the references in the last column).}
 \tablenotetext{e}{Is the PWN detected in radio?
P = possibly. }
 \tablenotetext{f}{References to the original papers where
the corresponding {\sl Chandra} and {XMM-Newton}
 data have been analyzed.  --
(1) Wang et al.\ 2001;
 (2) Chen et al.\ 2006; (3) Kaspi et al.\ 2001; (4) Gaensler et al.\ 2004;
(5) Petre et al.\ 2002; (6) McGowan et al.\ 2006;
(7) Becker et al.\ 2006; (8) Misanovic et al.\ 2007; tw = this work. }
\label{pulsar-tails}
\end{table*}

\end{document}